\documentclass[a4paper,11pt]{article}
\usepackage{jheppub}
\usepackage[numbers, sort&compress]{natbib}
\usepackage{amsmath, amsfonts, amssymb}
\usepackage{physics}
\usepackage{xcolor}
\usepackage{booktabs}
\usepackage{siunitx}
\usepackage{slashed}
\usepackage{orcidlink}
\allowdisplaybreaks

\title{
Revisiting lepton flavor violation: $\tau$ and meson decays
}
\author{Kevin A.\ Urqu\'ia-Calder\'on\orcidlink{0000-0002-7345-3428},}
\author{and Oleg Ruchayskiy\orcidlink{0000-0001-8073-3068}}

\affiliation{Niels Bohr Institute, University of Copenhagen, Jagtvej~155A, DK-2200, Copenhagen, Denmark}
\emailAdd{kevin.urquia@nbi.ku.dk}
\emailAdd{oleg.ruchayskiy@nbi.ku.dk}

\abstract{
The minimal type-I seesaw model provides a simple explanation of neutrino flavor oscillations and induces charged lepton flavor violation (cLFV). 
Despite extensive previous studies, semileptonic cLFV channels remain underexplored. 
Using updated form factors, decay constants, and oscillation data, we revisit $\tau$ and meson decay channels, performing a systematic comparison across the seesaw parameter space. 
Surprisingly, we find that decays such as $\tau \to \ell\rho$ can dominate over purely leptonic $\tau$-sector probes, including $\tau\to 3\ell$ and even $\tau\to\ell\,\gamma$, in certain regions. 
In contrast, heavy-meson decays remain far below experimental sensitivity. 
Considering global constraints on the seesaw parameters, we derive branching ratios for the relevant cLFV processes and identify those within potential reach of next-generation experiments.
}

\begin{document}
\maketitle

\section{Introduction}
The discovery of neutrino oscillations in Super-Kamiokande \cite{Super-Kamiokande:1998kpq} gave us the first laboratory evidence of physics beyond the Standard Model (SM). 
Neutrino oscillations imply that there must be mixing in the lepton sector and that neutrinos have masses. The SM predicts none.
A minimal extension to account for neutrino oscillations is by adding Heavy Neutral Leptons (HNLs, or also known as right-handed neutrinos or sterile neutrinos), which eventually give neutrino masses via the seesaw mechanism \cite{Minkowski:1977sc, Yanagida:1979as, Gell-Mann:1979vob, Mohapatra:1979ia, Schechter:1980gr, Schechter:1981cv}.
The presence of HNLs in the SM induces mixing in the lepton sector, which not only can explain neutrino oscillations, but it can also induce rare processes like charged lepton flavor violating processes (cLFV) \cite{Lee:1977qz, Petcov:1976ff, Lee:1977tib, Minkowski:1977sc, Bilenky:1977du, Marciano:1977wx, Cheng:1980tp, Lim:1981kv, Langacker:1988up, Pilaftsis:1992st, Ilakovac:1994kj, Illana:1999ww, Ilakovac:1999md, Illana:2000ic, Pascoli:2003rq, Pascoli:2003uh, Arganda:2004bz, Ibarra:2011xn, Dinh:2012bp, Alonso:2012ji, Hernandez-Tome:2020lmh, Crivellin:2022cve}.

cLFV channels such as $\mu\to e\gamma, \mu\to 3e$, and $\mu - e$ conversion in a nucleus have been systematically studied before, and have the strongest constraints. 
Many $\tau$ semileptonic channels (e.g. $\tau \to \ell_\alpha\,P$ and $\tau \to \ell_\alpha\,V$) have received much less attention within this model (however several studies have been done in the effective field theory perspective, see e.g. \cite{Petrov:2013vka, Celis:2013xja, Celis:2014asa, Hazard:2016fnc, Hazard:2017udp, Husek:2020fru, Angelescu:2020uug, Cirigliano:2021img, Calibbi:2022ddo, Descotes-Genon:2023pen, Fernandez-Martinez:2024bxg}).
The last systematic analyses done with these semileptonic $\tau$ decays in this model were done more than thirty years ago \cite{Ilakovac:1995km, Ilakovac:1995wc}.
Meson cLFV decays, on the other hand, have also not received much attention since \cite{Fajfer:1998px} (however, see \cite{Awasthi:2024nvi} for a recent study on a set of processes we will not consider in this paper).\footnote{Meson decays have been used in direct HNL searches, in different processes such as $P\to\ell N$, and indirectly from measurements of deviation from lepton universality \cite{Shrock:1980vy,Shrock:1980ct,Bondarenko:2018ptm,Bryman:2019bjg}.}

Our knowledge of form factors and meson decay constants, as well as oscillation data \cite{deSalas:2020pgw, Esteban:2024eli, Capozzi:2025wyn} has greatly improved since these studies were made.
Moreover, upcoming facilities (Belle-II \cite{Belle-II:2022cgf}, BES-III \cite{BESIII:2020nme}, STCF \cite{Achasov:2023gey}, REDTOP \cite{Gan:2020aco, REDTOP:2022slw}) aim to provide strong constraints on several cLFV processes.

Therefore, it is timely to revisit these processes.

In this paper, we re-derive branching ratio formulas for $\tau$ and meson decays in the minimal seesaw model. 
We provide indirect bounds, predictions, and a comparative analysis across channels. 
Our main results are: (a) $\tau \to \ell_\alpha\,\rho$ gives stronger bounds than $\tau \to 3\ell_\alpha$, and $\tau \to \ell_\alpha\,\gamma$ in some regions, and that $\tau \to \ell_\alpha\,\pi$ can also be competitive;
(b) light unflavored meson and quarkonia decays are much smaller, far below any experimental sensitivity; and (c) the maximum value of the branching ratio $\tau\to \ell_\alpha \rho$ is close to the sensitivity of Belle-II.

This paper is structured as follows: Section~\ref{sec:theory} introduces the theoretical framework and our notations. 
Section~\ref{sec:formulas} presents our formulas for different processes and systematically analyzes them.
Section~\ref{sec:predictions} gives indirect bounds to all analyzed processes.
Section~\ref{sec:bounds} shows bounds and compares our processes with more canonical decays. 
Finally, in Section~\ref{sec:conclusions} we summarize and conclude.

\section{Type-I seesaw theory} \label{sec:theory}
We extend the SM by including $\mathcal{N}$ neutral singlet right-handed fermions, $N_R$. 
Having no SM quantum numbers, their Lagrangian terms are
\begin{align}
    \label{eq:type_i_seesaw_lagrangian}
    \mathcal{L} = \mathcal{L}_\mathrm{SM} + i \bar{N}_R \slashed{\partial} N_R - \bar{L}_L \cdot \tilde{H}\,\mathbf{Y}_\nu N_R - \frac{1}{2}\bar{N}_R^C\,\mathbf{M}_M 
    \, N_R + \mathrm{H.c.}\,,
\end{align}
where $L_L = \begin{pmatrix}
    \nu_L \\ \ell_L
\end{pmatrix}$ is the SM fermion doublet, $\tilde{H} = i\sigma_2 H^\ast$ where $H = \begin{pmatrix}
    \varphi_W^+ \\ \frac{1}{\sqrt{2}}(H + i \varphi_Z^0)
\end{pmatrix}$, is the SM Higgs doublet. $\mathbf{Y}_\nu$ is a Yukawa matrix of dimensions $3 \times \mathcal{N}$, and $\mathbf{M}_M$ is a Majorana mass matrix of dimension $\mathcal{N}$.

After spontaneous symmetry breaking (SSB), the Higgs field acquires a vacuum expectation value (VEV), $v$, giving tree-level masses to SM particles, and now also providing a mixing term between the neutral singlets and active neutrinos. The mixing term looks like
\begin{align}
    \label{eq:mass_lagrangian}
    \mathcal{L}_\mathrm{mass} = -\frac{1}{2} \begin{pmatrix}
        \bar{\nu}_L & \bar{N}_R^C 
    \end{pmatrix} \begin{pmatrix}
        0 & \mathbf{M}_D \\ \mathbf{M}_D^T & \mathbf{M}_M
    \end{pmatrix} \begin{pmatrix}
        \nu_L^C \\ N_R
    \end{pmatrix}\,,
\end{align}
where $\mathbf{M}_D = \frac{1}{\sqrt{2}}v\mathbf{Y}_\nu$. 
The diagonalization into mass eigenstates provides us with the two following states
\begin{align}
    n_L &\simeq \left(\mathbf{I} - \frac{1}{2}\mathbf{\Theta}\,\mathbf{\Theta}^\dagger\right)\mathbf{V}\,\nu_L + \mathbf{\Theta}\,N_R^C\,, &
    \tilde{n}_R &\simeq -\mathbf{\Theta}^T\,\mathbf{V}^\ast \nu_L^C + \left(\mathbf{I} - \frac{1}{2} \mathbf{\Theta}^T\,\mathbf{\Theta} \right) N_R\,,
\end{align}
where $\mathbf{V}$ is the usual unitary PMNS matrix, and $\mathbf{\Theta} \simeq -\mathbf{M}_D\,\mathbf{M}_M^{-1}$ is the mixing angle matrix between the singlets and active neutrinos. 
We expanded only up to $\mathcal{O}\left(\mathbf{\Theta}^2\right)$. 
This expansion is justified because we know from experimental data and precision measurements, that matrix elements must abide $\mathbf{\Theta} \lesssim \num{e-3}$.

The diagonalization of this matrix gives the mass spectrum of neutral fermions. 
In the limit where $\mathbf{M}_D \ll \mathbf{M}_M$, we arrive at the following relation for active neutrino masses
\begin{align}
    \label{eq:seesaw}
    \mathbf{V}\,\mathbf{m}_\nu\,\mathbf{V}^T &\simeq - \mathbf{M}_D\, \mathbf{M}_N^{-1}\,\mathbf{M}_D^T\,,&
    \mathbf{M}_N &\simeq \mathbf{M}_M\,, \\
    &\simeq -\mathbf{\Theta}\,\mathbf{M}_N\,\mathbf{\Theta}^T\,. \nonumber
\end{align}
This is the \textit{type-I seesaw relation}. 
We should also highlight here that we considered the first power of $\mathbf{\Theta}$ in this relation. 

Interactions between neutral leptons and the rest of the SM particles, in the Feynman-'t-Hooft gauge, are
\begin{align}
    \mathcal{L}_{W^\pm} &= \frac{g}{\sqrt{2}}\,\mathcal{U}_{\alpha i}\,\bar{\ell}_\alpha\,\slashed{W}^-\,P_L\,n_i + \mathrm{h.c.}\,, \\
    \mathcal{L}_Z &= \frac{g}{2\,c_w}\,\bar{n}_i\,\slashed{Z}\left[\mathcal{C}_{ij} P_L - \mathcal{C}^\ast_{ij} P_R\right]\,n_j\,, \\
    \label{eq:charged_goldstone_interactions}
    \mathcal{L}_{\phi^\pm} &= \frac{g}{\sqrt{2}\, M_W}\,\mathcal{U}_{\alpha i}\,\phi^-\,\bar{\ell}_\alpha \left[m_i\,P_R - m_\alpha\,P_L\right] n_i + \mathrm{h.c.} \,, \\
    \label{eq:neutral_goldstone_interactions}
    \mathcal{L}_{\phi^0} &= \frac{i\,g}{4\,M_W}\,\phi^0\, \bar{n}_i \left[\mathcal{C}_{ij} (m_j\,P_R - m_i\,P_L) + \mathcal{C}_{ij}^{\ast} (m_i\,P_R - m_j\,P_L)\right]\,n_j\,, \\
    \label{eq:higgs_interactions}
    \mathcal{L}_{H} &= -\frac{g}{4\,M_W}\,H\, \bar{n}_i \left[\mathcal{C}_{ij} (m_j\,P_R + m_i\,P_L) + \mathcal{C}_{ij}^{\ast} (m_i\,P_R + m_j\,P_L)\right]\,n_j\,,
\end{align}
where we introduced the matrices $\mathcal{U} \equiv \left[
    (\mathbf{I} - \frac{1}{2}\mathbf{\Theta}\mathbf{\Theta}^\dagger)\,\mathbf{V},\mathbf{\Theta}
\right], \mathcal{C} \equiv \mathcal{U}^\dagger \mathcal{U}$. In Appendix~\ref{app:unitarity_seesaw_formulas} we summarize some important relationships regarding the unitarity of $\mathcal{U}$ and $\mathcal{C}$.

In what remains of the paper, we will solely focus on the case where $\mathcal{N} = 2$. 
This is the most minimal case that can explain neutrino oscillation data, but it also predicts that the lightest neutrino is massless. 

\subsection{Parametrization of mixing angles in the minimal model with 2 HNLs}

In the vanilla realization of the seesaw mechanism, the smallness of the active neutrino masses is attributed to the largeness of $\mathbf{M}_N$ compared to that of $\mathbf{M}_D$. 
This \textit{vanilla} realization of the seesaw model is very hard to probe, since it would also give us extremely small $\mathbf{\Theta}$ that almost no experiment can probe.

However, we can still have large mixing angles and radiatively small neutrino masses.
We can do this by imposing that the smallness of neutrinos is generated by the small breaking of a symmetry. 
In the minimal model with only two additional HNLs, we can achieve this by imposing $M_{N1} = M_{N2} = M_N$ and $\Theta_{\alpha 1} = \pm i\,\Theta_{\alpha 2} = \frac{1}{\sqrt{2}} \Theta_{\alpha}$.
This will leave neutrinos massless at tree-level. 
Thus, neutrino masses are generated from small perturbations to the equalities we just described \cite{Shaposhnikov:2006nn, Kersten:2007vk, Abada:2007ux, Adhikari:2010yt, Ibarra:2010xw, Moffat:2017feq}. 

To properly incorporate neutrino oscillation data into our analyses with large mixing angles, we use the Casas-Ibarra parametrization \cite{Casas:2001sr}.
Solving for $\mathbf{\Theta}$ in Eq.~\eqref{eq:seesaw}
\begin{align}
    \mathbf{\Theta} = i\,\mathbf{V}\mathbf{m_\nu}^{1/2}\,\mathbf{\Omega}\,\mathbf{M}_N^{-1/2}\,,
\end{align}
where $\mathbf{\Omega}$ is an arbitrary semi-orthogonal matrix that is only parametrized by one angle. 
This matrix differs for the normal mass ordering (NO) and for the inverted mass ordering (IO) of neutrino masses
\begin{align}
    \mathbf{\Omega}_{\mathrm{NO}} &= \begin{pmatrix}
        0 & 0 \\
        \cos\omega & \sin\omega \\
        -\sin\omega & \cos\omega
    \end{pmatrix}\,,&
    \mathbf{\Omega}_{\mathrm{IO}} &= \begin{pmatrix}
        \cos\omega & \sin\omega \\
        -\sin\omega & \cos\omega \\
        0 & 0
    \end{pmatrix}\,,
\end{align}
the $L$ conserved scenario is obtained in the limit where $\Im(\omega) \gg 0$. 
In this limit, the Casas-Ibarra matrices become
\begin{align}
    \mathbf{\Omega}_{\mathrm{NO}} &\approx \frac{e^{-i\omega}}{2}\begin{pmatrix}
        0 & 0 \\
        1 & -i \\
        i & 1
    \end{pmatrix}\,,&
    \mathbf{\Omega}_{\mathrm{IO}} &\approx 
    \frac{e^{-i\omega}}{2}\begin{pmatrix}
        1 & -i \\
        i & 1 \\
        0 & 0
    \end{pmatrix}\,.
\end{align}

We can write the total mixing angle $U_{\mathrm{tot}}^2 = \sum_{\alpha i} \abs{\Theta_{\alpha i}}^2 = \sum_{\alpha} \abs{\Theta_{\alpha}}^2$, in terms of $\omega, M_N$ and neutrino masses
\begin{align}
    U_{\mathrm{tot}}^2 = e^{2\Im(\omega)}\frac{\sum m_\nu}{2\,M_N}\,.
\end{align}

Now we introduce
\begin{align}
    x_\alpha \equiv \frac{\sum_{i = 1}^2 \abs{\Theta_{\alpha i}}^2}{U_{\mathrm{tot}}^2} = \frac{\abs{\Theta_{\alpha}}^2}{U_{\mathrm{tot}}^2} \,,
\end{align}
these quantities are written only in terms of neutrino oscillation data and a free Majorana phase \cite{Ruchayskiy:2011aa, Ibarra:2011xn}
\begin{align}
    \label{eq:x_alpha_definition}
    x_\alpha = \frac{\abs{V_{\alpha i}\sqrt{m_{\nu,i}} + i\,V_{\alpha j} \sqrt{m_{\nu,j}} }^2}{\sum_k m_{\nu,k}}\,.
\end{align}

\begin{figure}[!t]
    \centering
    \includegraphics[width=0.75\linewidth]{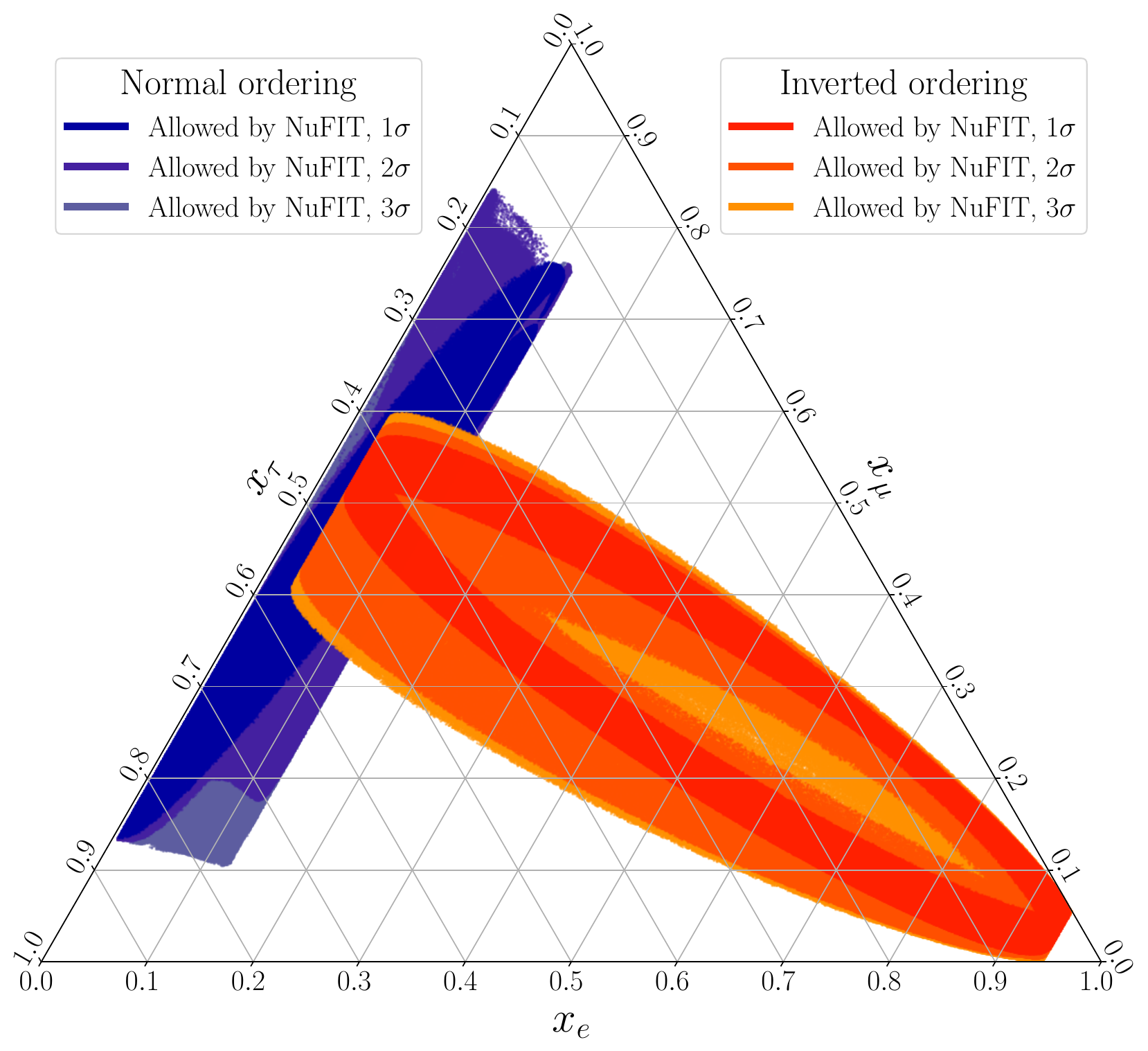}
    \caption{Ternary plot showing the allowed values of $x_\alpha$ consistent with neutrino oscillation data. 
    The plot was done by letting $\delta$ and $\theta_{23}$ run free, and choosing the allowed regions of parameter space according to the values tabulated by NuFit 6.0. \cite{Esteban:2024eli}.}
    \label{fig:flavor_triangle}
\end{figure}
\begin{table}[!t]
    \centering
    \begin{tabular}{r|r}
         Normal ordering & Inverted ordering  \\ \hline
          $\num{0.004} < x_e < \num{0.125}$ 
          & $\num{0.023} < x_e < \num{0.944}$  \\  
         $\num{0.123} < x_\mu < \num{0.862}$ 
         & $\num{0} < x_\mu < \num{0.594}$  \\
         $\num{0.105} < x_\tau < \num{0.840}$ 
         & $\num{0} < x_\tau < \num{0.609}$
    \end{tabular}\hfill
    \begin{tabular}{r|r}
         Normal ordering & Inverted ordering  \\ \hline
          $\num{0.027} < \abs{\Lambda_{e\mu}} < \num{0.308}$ &
          $\num{0} < \abs{\Lambda_{e\mu}} < \num{0.425}$  \\ 
         $\num{0.025} < \abs{\Lambda_{e\tau}} < \num{0.305}$ &
         $\num{0} < \abs{\Lambda_{e\tau}} < \num{0.433}$  \\
         $\num{0.285} < \abs{\Lambda_{\mu\tau}} < \num{0.484}$ &
         $\num{0} < \abs{\Lambda_{\mu\tau}} < \num{0.489}$
    \end{tabular}
    \caption{Allowed values of $x_{\alpha}$ and $\Lambda_{\alpha\beta}$ from neutrino oscillation data. The values were obtained with the global analyzes performed by NuFIT 6.0. The bounds did not significantly change when comparing the allowed by the $1\sigma, 2\sigma$ and $3\sigma$ regions.}
    \label{tab:allowed_ratios}
\end{table}

In the context of cLFV processes, it is also pertinent to identify the product from two different angles. 
In particular, it will also define
\begin{align}
    \Lambda_{\alpha\beta} \equiv  \frac{\sum_{i} \Theta_{\alpha i}^{\phantom{\ast}}\,\Theta_{\beta i}^{\ast}}{U_{\mathrm{tot}}^2}\,.
\end{align}

We show the allowed values of each value of $x_\alpha$ in Fig.~\ref{fig:flavor_triangle} as allowed by neutrino oscillation data, from the recent global analysis performed by NuFIT 6.0 \cite{Esteban:2024eli}. 
We did the ternary plot by letting $\theta_{23}$ and $\delta$ free, and using the tabulated data from 2d plots of NuFIT as the allowed regions. 
We chose these two because they have the largest uncertainty among all neutrino oscillation data.
We also summarize the maximum and minimum values of each $x_{\alpha}$ and $\Lambda_{\alpha\beta}$ in Table~\ref{tab:allowed_ratios}.

The most salient feature of both Fig.~\ref{fig:flavor_triangle} and Table~\ref{tab:allowed_ratios} is the fact that neutrino oscillation data allows for either $x_\mu$ or $x_\tau$ to be zero for the inverted mass ordering. 
The conditions for this to happen were already highlighted in \cite{Ruchayskiy:2011aa, Ibarra:2011xn}, but we will summarize them now. 
In Eq.~\ref{eq:x_alpha_definition}, we need both that the complex phase of $V_{\alpha,i}$ and $V_{\alpha,j}$ to differ by $\pi/2$, and for $\abs{V_{\alpha,i}} \sqrt{m_{\nu,i}} = \abs{V_{\alpha,j}} \sqrt{m_{\nu,j}}$. 
The former can always be fulfilled, due to having an unconstrained Majorana phase. The latter, however, is only fulfilled for the IO for $\alpha = \mu$ and $\alpha = \tau$.  

A similar asymmetrical mixing pattern has been discussed before in \cite{Cottin:2022nwp}. 
Asymmetical mixing where a flavor dominates over others naturally arises in low-scale seesaw realizations.

\section{cLFV processes with mesons} \label{sec:formulas}
\begin{figure}
    \centering
    \includegraphics[width=\linewidth]{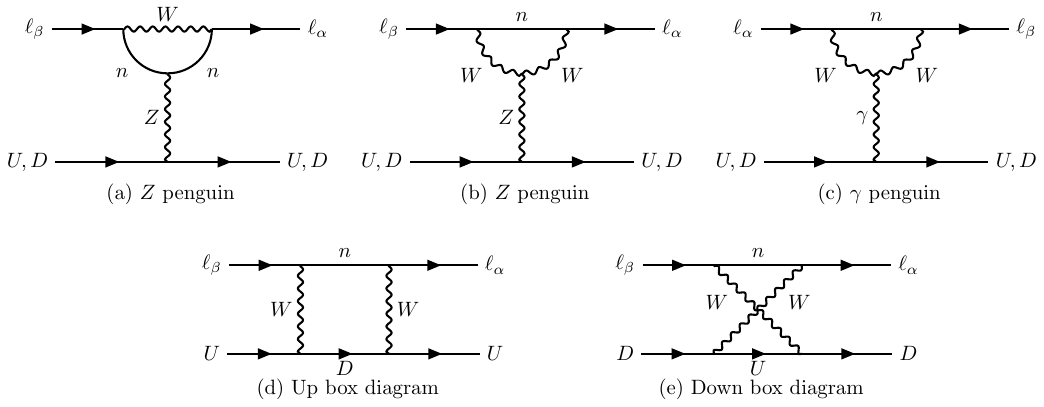}
    \caption{Feynman diagrams mediating cLFV processes with quarkonia.}
    \label{fig:diagrams}
\end{figure}

\begin{table}[t]
    \centering
    \resizebox{0.47\textwidth}{!}{%
    \begin{tabular}[t]{l|l|l}
    \toprule
    Process & Current bound & Future sensitivity \\
    \midrule 
    \multicolumn{3}{c}{Processes proportional to $\Lambda_{e\tau}$} \\
    \midrule
    $\tau \to e\gamma$ & \num{2.6e-8}~\cite{HeavyFlavorAveragingGroupHFLAV:2024ctg} & \num{9.0e-9}~\cite{Belle-II:2022cgf} \\ 
    $\tau \to 3e$ & \num{1.9e-8}~\cite{HeavyFlavorAveragingGroupHFLAV:2024ctg} & \num{4.7e-10}~\cite{Belle-II:2022cgf} \\
    \midrule
    $\tau \to e \pi$ & \num{6.4e-8e-8}~\cite{HeavyFlavorAveragingGroupHFLAV:2024ctg} & \num{7.3e-10}~\cite{Belle-II:2022cgf} \\
    $\tau \to e \eta$ & \num{7.4e-8}~\cite{HeavyFlavorAveragingGroupHFLAV:2024ctg} & \num{1.2e-9}~\cite{Belle-II:2022cgf} \\
    $\tau \to e \eta^\prime$ & \num{1.3e-7}~\cite{HeavyFlavorAveragingGroupHFLAV:2024ctg} & \num{1.2e-9}~\cite{Belle-II:2022cgf} \\
    \midrule
    $\tau \to e \rho$ & \num{2.0e-8}~\cite{HeavyFlavorAveragingGroupHFLAV:2024ctg} & \num{3.8e-10}~\cite{Belle-II:2022cgf} \\
    $\tau \to e \omega$ & \num{2.2e-8}~\cite{HeavyFlavorAveragingGroupHFLAV:2024ctg} & \num{1.0e-9}~\cite{Belle-II:2022cgf} \\
    $\tau \to e \phi$ & \num{1.6e-8}~\cite{HeavyFlavorAveragingGroupHFLAV:2024ctg} & \num{7.4e-10}~\cite{Belle-II:2022cgf} \\
    \midrule
    $\tau \to e\pi^+\pi^-$ & \num{2.1e-8}~\cite{HeavyFlavorAveragingGroupHFLAV:2024ctg} & \num{5.8e-10}~\cite{Belle-II:2022cgf} \\
    && $\mathcal{O}(\num{e-10})$~\cite{Achasov:2023gey}\\
    $\tau \to e K^+K^-$ & \num{3.3e-8}~\cite{HeavyFlavorAveragingGroupHFLAV:2024ctg} & \num{6.5e-10}~\cite{Belle-II:2022cgf} \\
    && $\mathcal{O}(\num{e-10})$~\cite{Achasov:2023gey}\\
    $\tau \to e K_S K_S$ & \num{7.1e-8}~\cite{Belle:2010rxj} & \num{9.7e-10}~\cite{Belle-II:2022cgf} \\
    \midrule 
    \multicolumn{3}{c}{Processes proportional to $\Lambda_{\mu\tau}$} \\
    \midrule
    $\tau \to \mu\gamma$ & \num{2.7e-8}~\cite{HeavyFlavorAveragingGroupHFLAV:2024ctg} & \num{9.0e-9}~\cite{Belle-II:2022cgf} \\
    $\tau \to 3\mu$ & \num{1.6e-8}~\cite{HeavyFlavorAveragingGroupHFLAV:2024ctg} & \num{3.6e-10}~\cite{Belle-II:2022cgf} \\
    \midrule
    $\tau \to \mu \pi$ & \num{8.1e-8}~\cite{HeavyFlavorAveragingGroupHFLAV:2024ctg} & \num{7.1e-10}~\cite{Belle-II:2022cgf} \\
    $\tau \to \mu \eta$ & \num{6.1e-8}~\cite{HeavyFlavorAveragingGroupHFLAV:2024ctg} & \num{8.0e-10}~\cite{Belle-II:2022cgf} \\
    $\tau \to \mu \eta^\prime$ & \num{8.6e-8}~\cite{HeavyFlavorAveragingGroupHFLAV:2024ctg} & \num{1.2e-9}~\cite{Belle-II:2022cgf} \\
    \midrule
    $\tau \to \mu \rho$ & \num{1.3e-8}~\cite{HeavyFlavorAveragingGroupHFLAV:2024ctg} & \num{5.5e-10}~\cite{Belle-II:2022cgf} \\
    $\tau \to \mu \omega$ & \num{3.4e-8}~\cite{HeavyFlavorAveragingGroupHFLAV:2024ctg} & \num{1.4e-9}~\cite{Belle-II:2022cgf} \\
    $\tau \to \mu \phi$ & \num{2.6e-8}~\cite{HeavyFlavorAveragingGroupHFLAV:2024ctg} & \num{8.4e-10}~\cite{Belle-II:2022cgf} \\
    \midrule
    $\tau \to \mu\pi^+\pi^-$ & \num{2.1e-8}~\cite{Belle:2012unr} & \num{5.6e-10}~\cite{Belle-II:2022cgf} \\
    && $\mathcal{O}(\num{e-10})$~\cite{Achasov:2023gey} \\
    $\tau \to \mu K^+K^-$ & \num{4.2e-8}~\cite{HeavyFlavorAveragingGroupHFLAV:2024ctg} & \num{1.1e-9}~\cite{Belle-II:2022cgf} \\
    && $\mathcal{O}(\num{e-10})$~\cite{Achasov:2023gey} \\
    $\tau \to \mu K_S K_S$ & \num{8.0e-8}~\cite{Belle:2010rxj} & \num{1.1e-9}~\cite{Belle-II:2022cgf} \\ \midrule
    \end{tabular}}\hfill%
    \resizebox{0.47\textwidth}{!}{%
    \begin{tabular}[t]{l|l|l}
    \toprule
    Process & Current bound & Future sensitivity \\
    \midrule 
    \multicolumn{3}{c}{Light meson decays} \\
    \midrule
    $\pi \to e\mu$ & \num{3.2e-10}~\cite{NA62:2021zxl} &  \\ 
    $\eta \to e\mu$ & \num{6e-6}~\cite{White:1995jc} & \\
    $\eta^\prime \to e\mu$ & \num{4.7e-4}~\cite{CLEO:1999nsy} &  \\ \midrule
    $\rho \to e \mu$ & & \\ 
    $\omega \to e \mu$ & &  \\
    $\phi \to e \mu$ & \num{2.0e-6}~\cite{Achasov:2009en} & \\ \midrule
    \multicolumn{3}{c}{Quarkonia decays}
    \\ \midrule
    $J/\Psi \to e\mu$ & \num{1.6e-7}~\cite{BESIII:2013jau}& \num{6.0e-9}~\cite{BESIII:2020nme} \\
    $J/\Psi \to e\tau$ & \num{8.3e-6}~\cite{BESIII:2021slj} & \num{2.5e-8}~\cite{BESIII:2020nme} \\
    $J/\Psi \to \mu\tau$ & \num{2.0e-6}~\cite{BES:2004jiw} & \num{1.5e-8}~\cite{BESIII:2020nme} \\ \midrule
    $\Upsilon \to e\mu$ & \num{3.9e-7}~\cite{Belle:2022cce} & \\
    $\Upsilon \to e\tau$ & \num{2.7e-7}~\cite{Belle:2022cce} & \\
    $\Upsilon \to \mu\tau$ & \num{2.7e-6}~\cite{Belle:2022cce} & \\ \midrule
    \end{tabular}}
    \caption{Current bounds and prospective sensitivity of different cLFV processes we will analyze in this paper. 
    The null entries indicate either no current bound or no planned searches.}
    \label{tab:cLFV_decays_limits}
\end{table}

cLFV processes are due to different loop diagrams, including boxes, penguins, and transitions, all of which are well-known in the literature. These diagrams are shown in Fig.~\ref{fig:diagrams}.

We list the processes we aim to study in Table~\ref{tab:cLFV_decays_limits}.
The bounds from $\tau$ decays do not match the ones listed by the PDG because we chose to use the ones by the Heavy Flavor Averaging Group (HFLAV) \cite{HeavyFlavorAveragingGroupHFLAV:2024ctg}, that does a combined analysis of the bounds made by Belle and BaBar. 
The null entries in Table~\ref{tab:cLFV_decays_limits} indicate there is no current bound, or no prospective sensitivity listed. 
The Jefferson Eta Factory (JEF) \cite{Mack:2014gma, Somov:2024jiy} and REDTOP \cite{Gan:2020aco, REDTOP:2022slw} future experiments aim to place strong bounds on $\pi, \eta, \eta^\prime$ cLFV decays, but there is no estimate on their sensitivity.

The amplitudes we need to compute all these processes are
\begin{align}
    \label{eq:monopole_amplitude}
    i\mathcal{M}_\gamma &= 
        i\frac{\alpha_W^2\,s_W^2}{2\,M_W^2} 
            \left[\bar{q} \gamma^\mu \mathcal{Q}_q q \right] 
            \left[\bar{\ell}_\beta F_\gamma^{\beta\alpha} 
                \left(\gamma_\mu - \frac{\slashed{q}\,q_\mu}{q^2} \right) 
            P_L \ell_\alpha \right]\,, \\
    \label{eq:dipole_amplitude}
    i\mathcal{M}_{\gamma-\mathrm{dipole}} &= 
        i\frac{\alpha_W^2\,s_W^2}{2\,M_W^2} 
            \left[\bar{q} \gamma^\mu \mathcal{Q}_q q \right] 
            \left[\bar{\ell}_\beta 
                \left(-i\frac{\sigma_{\mu\nu} q^\nu}{q^2}
                \left(m_\beta P_L + m_\alpha P_R\right)\right) 
            G_\gamma^{\beta\alpha} \ell_\alpha \right]\,, \\
    \label{eq:Z_amplitude}
    i\mathcal{M}_Z &= 
        i\frac{\alpha_W^2}{2\,M_W^2} 
            \left[\bar{q} \gamma^\mu \left(\mathcal{T}_L^3\,P_L - s_W^2 \mathcal{Q}_q\right) q \right] 
            \left[\bar{\ell}_\beta F_Z^{\beta\alpha} \gamma_\mu P_L \ell_\alpha \right]\,, \\
    \label{eq:up_box_amplitude}
    i\mathcal{M}_{\mathrm{box}}^{u_i u_j} &= 
        i\frac{\alpha_W^2}{4\,M_W^2} 
            \left[\bar{U}_j \gamma^\mu P_L U_i \right] 
                \left[\bar{\ell}_\beta F_{\mathrm{u,box}}^{ji\beta\alpha} \gamma_\mu P_L \ell_\alpha \right]\,, \\
    \label{eq:down_box_amplitude}
    i\mathcal{M}_{\mathrm{box}}^{d_j d_i} &= 
        -i\frac{\alpha_W^2}{4\,M_W^2} 
            \left[\bar{D}_i \gamma^\mu P_L D_j \right] 
            \left[\bar{\ell}_\beta F_{\mathrm{d,box}}^{ji\beta\alpha} \gamma_\mu P_L \ell_\alpha \right]\,.
\end{align}
where $\alpha_W = \frac{g^2}{4\pi}$, where $g$ is the weak coupling, $q$ is the momentum of the photon that mediates interactions between the photons and quarks. The subindices $i,j$ can represent different quark flavors. All the loop functions above, $F_{\gamma}^{\beta\alpha}, G_{\gamma}^{\beta\alpha}, \dots$, are functions of HNL parameters, including mixing angles and HNL masses. They are defined in Appendix~\ref{app:definitions_f_and_g}.

The amplitudes in Eqs.~\eqref{eq:monopole_amplitude} -- \eqref{eq:Z_amplitude} respect flavor conservation in the quark sector. 
However, the amplitudes in Eqs.~\eqref{eq:up_box_amplitude} and \eqref{eq:down_box_amplitude} may induce flavor violation in the quark sector; thus, we could have cLFV processes with flavored mesons, but we would expect these to be suppressed due to the GIM mechanism \cite{Glashow:1970gm}.

\begin{table}[t!]
    \centering
    \begin{tabular}{c|c|c}
        Initial/final state & Current & Form factor/decay width  \\ \toprule
        $\ket{\pi^0}$ & $\frac{1}{\sqrt{2}}(\bar{u}\gamma^\mu\gamma_5 u  - \bar{d}\gamma^\mu\gamma_5 d)$ & $f_{\pi^0} = \SI{130.2}{MeV}$ \cite{FlavourLatticeAveragingGroupFLAG:2024oxs} \\
        $\ket{\eta_8}$ & $\frac{1}{\sqrt{6}}(\bar{u}\gamma^\mu\gamma_5 u  + \bar{d}\gamma^\mu\gamma_5 d - 2 \bar{s}\gamma^\mu\gamma_5 s)$ & $f_8 = \SI{115.0}{MeV}$ \cite{Bali:2021qem} \\ 
        $\ket{\eta_0}$ & $\frac{1}{\sqrt{3}}(\bar{u}\gamma^\mu\gamma_5 u  + \bar{d}\gamma^\mu\gamma_5 d + \bar{s}\gamma^\mu\gamma_5 s)$ & $f_0 = \SI{100.1}{MeV}$ \cite{Bali:2021qem} \\
        $\ket{\rho}$ & $\frac{1}{\sqrt{2}}(\bar{u}\gamma^\mu u  - \bar{d}\gamma^\mu d)$ & $f_{\rho} = \SI{213}{MeV}$ \cite{Bharucha:2015bzk}  \\
        $\ket{\omega}$ & $\frac{1}{\sqrt{2}}(\bar{u}\gamma^\mu u  + \bar{d}\gamma^\mu d)$ &  $f_{\omega} = \SI{197}{MeV}$ \cite{Bharucha:2015bzk}   \\
        $\ket{\phi}$ & $\bar{s}\gamma^\mu s$ & $f_{\phi} = \SI{233}{MeV}$ \cite{Bharucha:2015bzk} \\
        $\ket{J/\psi}$ & $\bar{c}\gamma^\mu c$ & $f_{J/\psi} = \SI{418}{MeV}$ \cite{Becirevic:2013bsa} \\
        $\ket{\Upsilon}$ & $\bar{b}\gamma^\mu b$ & $f_\Upsilon = \SI{649}{MeV}$ \cite{Colquhoun:2014ica} \\
        $\ket{\pi^+ \pi^-}$ & $\frac{1}{2}(\bar{u}\gamma^\mu u  - \bar{d}\gamma^\mu d)$ & $F_{\pi\pi}(s)$ \\
        $\ket{K K}$ & $\frac{1}{2}(\bar{u}\gamma^\mu u  - \bar{d}\gamma^\mu d)$ & $F_{KK}^{(3)}(s)$ \\
        $\ket{K K}$ & $\frac{1}{6}(\bar{u}\gamma^\mu u  + \bar{d}\gamma^\mu d - 2 \bar{s}\gamma^\mu s)$ & $F_{KK}^{(8)}(s)$ \\
        $\ket{K K}$ & $\frac{1}{3}(\bar{u}\gamma^\mu u  + \bar{d}\gamma^\mu d + \bar{s}\gamma^\mu s)$ & $F_{KK}^{(0)}(s)$ \\
        \midrule 
        $\ket{\eta}$ & $c_8 \ket{\eta_8} - s_0 \ket{\eta_0}$ & $\theta_8 = \ang{-25.8}$ \cite{Bali:2021qem} \\
        $\ket{\eta^\prime}$ & $s_8 \ket{\eta_8} + c_0 \ket{\eta_0}$ & $\theta_0 = \ang{-8.1}$ \cite{Bali:2021qem} \\
        $\ket{K^+ K^-}$ & $\ket{KK}_3 + \ket{KK}_8 + \ket{KK}_0$ \\
        $\ket{K^0 \bar{K}^0}$ & $-\ket{KK}_3 + \ket{KK}_8 + \ket{KK}_0$ \\ \midrule
    \end{tabular}
    \caption{Quark composition of different mesons. For the definition of the energy-dependent form factors, see Appendix~\ref{app:form_factors}.}
    \label{tab:bound_states_mesons}
\end{table}

To obtain the branching ratios of cLFV processes involving pseudoscalar (which we will denote by $P$) or vector ($V$) mesons, we will use the following conventions for their form factors
\begin{align}
    \mel{P(k)}{J_{5}^\mu}{0} &= -i f_P\,k^\mu\,,&
    \mel{0}{J_{5}^\mu}{P(k)} &= i f_P\,k^\mu\,,& \\
    \mel{V(k)}{J_{V}^\mu}{0} &= f_V m_V\,\epsilon_V^{\ast\mu}(k)\,,&
    \mel{0}{J_{V}^\mu}{V(k)} &= f_V m_V\,\epsilon_V^{\mu}(k)\,, \\
    \mel{P(k_1) P(k_2)}{J_{V}^\mu}{0} &= (k_1 - k_2)_\mu\,F_{PP}(s)\,, \label{eq:form_factor_from_amplitude}
\end{align}
where $J_5^\mu = \bar{q}_1 \gamma^\mu\,\gamma_5 q_2, J_V^\mu = \bar{q}_1 \gamma^\mu q_2$, where $q_1, q_2$ are the constituent quarks for each respective meson; in some cases, this should be a linear combination of different currents. $k$ is the momentum of the initial or final meson; $k_1$ and $k_2$ are the final momentum of the pair of pseudoscalar mesons. 
We also introduce $f_P$, pseudoscalar decay constants, $f_V$, the vector ones, and $F_{PP}(s)$, the energy-dependent form factors, where $s = (k_1 + k_2)^2$ is the invariant mass of the pseudoscalar system. 

We will not consider any processes with scalar mesons, such as $f_0$ or $a_0$. cLFV decays of these scalars are possible through Higgs or $Z$-boson channels.
These mesons are understudied, so their decay constants are unknown; moreover, we expect their branching ratios to cLFV to be very small. 

We define the currents, quark composition, decay constants, and form factor for each initial or final state meson in Table~\ref{tab:bound_states_mesons}. 

The shape of the amplitude of any individual process will heavily depend on the quark structure of each individual meson, as is shown in Table~\ref{tab:bound_states_mesons}. 
But the spin structure will be similar, depending on whether we have a pseudoscalar, two pseudoscalars, or one vector meson in the final state.
Pseudoscalar mesons do not interact with photons, but a vector or a pair of pseudoscalars will, and therefore they will receive contributions from diagrams where we have an on-shell (the \textit{dipole} term) or an off-shell photon (the \textit{monopole} term). The amplitudes for a $\ell_\beta \to \ell_\alpha (P, V, PP)$ are
\begin{align}
    \label{eq:pseudoscalar_amplitude}
    i\mathcal{M}(\ell_\beta \to \ell_\alpha P) &= \frac{\alpha_W^2}{2 M_W^2} p_P^\mu\,\mathcal{F}^{\beta\alpha}_P\,\left[\bar{\ell}_\beta\,\gamma^\mu P_L \ell_\alpha\right]\,, \\
    \label{eq:vector_amplitude}
    i\mathcal{M}(\ell_\beta \to \ell_\alpha V) &= 
    \begin{aligned}[t]
    &\frac{i\,\alpha_W^2}{2 M_W^2} \epsilon^{\ast\mu}_V(p_V)
    \biggl\{\mathcal{F}^{\beta\alpha}_{V} \left[\bar{\ell}_\beta\, \gamma^\mu P_L \ell_\alpha\right]  \\
    &+\mathcal{G}^{\beta\alpha}_{V} \left.\left[\bar{\ell}_\beta  \left(-i \frac{\sigma_{\mu\nu}\,p_V^\mu}{m_V^2}\right)(m_\beta P_L + m_\alpha P_R)\,\ell_\alpha \right]\right\}\,,
    \end{aligned} \\
    \label{eq:two_meson_amplitude}
    i\mathcal{M}(\ell_\beta \to \ell_\alpha P P) &= 
    \begin{aligned}[t]
    &\frac{i\,\alpha_W^2}{2 M_W^2} (p_{P1}^\mu - p_{P2}^\mu)
    \biggl\{\mathcal{F}^{\beta\alpha}_{PP} \left[\bar{\ell}_\beta\, \gamma^\mu P_L \ell_\alpha\right]  \\
    &+\mathcal{G}^{\beta\alpha}_{PP} \left.\left[\bar{\ell}_\beta  \left(-i \frac{\sigma_{\mu\nu}\,(p_{P1}^\nu + p_{P2}^\nu)}{s}\right)\,(m_\beta P_L + m_\alpha P_R)\,\ell_\alpha \right]\right\}\,,
    \end{aligned}
\end{align}
where the $\mathcal{F}$ and $\mathcal{G}$ are functions that have both the form factors, decay widths, and loop functions for each final state. All of these functions are defined in Appendix~\ref{app:defintions_of_F_and_G}. 
The shape for the amplitudes of meson decays, $P \to \ell_\alpha \ell_\beta$ and $V \to \ell_\alpha \ell_\beta$ can easily be derived from Eqs.~\eqref{eq:pseudoscalar_amplitude} and \eqref{eq:vector_amplitude}.
For the pseudoscalar amplitude, the amplitude will be negative, and for the vector amplitude, we would have the change $\epsilon_V^\ast \to \epsilon_V$.
We omit the term proportional to $\slashed{q}\,q^\mu$ from the photon contribution in Eq.~\eqref{eq:monopole_amplitude} as it vanishes when computing average amplitude squared, and is not relevant for any phenomenological analysis.

\subsection{Branching ratios}
We shall now write the branching ratios in the limit where the final leptons (an electron or muon) have negligible masses. The general equations are written in Appendix~\ref{app:branching_ratios_general}.

In this limit, and in terms of the mass ratios $x_P \equiv m_P^2/m_\tau^2, x_V \equiv m_V^2/m_\tau^2$, the formulas for branching ratios for $\tau$ decays are
\begin{align}
    \mathrm{Br}(\tau \to \ell_\alpha P) &= \frac{\alpha_W^4}{128\pi} \frac{m_\tau^4}{M_W^4} \frac{m_\tau}{\Gamma_\tau} \cdot \frac{\abs{\mathcal{F}_P^{\tau\alpha}}^2}{m_\tau^2}\left(1 - x_P\right)^2\,, \\[1.5ex]
    \mathrm{Br}(\tau \to \ell_\alpha V) &= \frac{\alpha_W^4}{128\pi} \frac{m_\tau^4}{M_W^4} \frac{m_\tau}{\Gamma_\tau} \cdot
    \left[
    \frac{\abs{\mathcal{F}_V^{\tau\alpha}}^2}{m_\tau^4}
    \left(\frac{1 + x_V - 2\,x_V^2}{x_V} \right) \right.\\ \nonumber
    & \quad + \frac{6\,\Re[\mathcal{F}_V^{\tau\alpha}\mathcal{G}_V^{\tau\alpha\ast}]}{m_\tau^4} 
    \left(\frac{1 - x_V}{x_V} \right) +  \left. \frac{\abs{\mathcal{G}_V^{\tau\alpha}}^2}{m_\tau^4}
    \left( \frac{2x_V^{-1} - x_V - 1}{x_V} \right) \right] \left(1 - x_V\right)\,,\\[1.5ex]
    \mathrm{Br}(\tau \to \ell_\alpha P P) &= \frac{\alpha_W^4}{128\pi} \frac{m_\tau^4}{M_W^4} \frac{m_\tau}{\Gamma_\tau} \cdot \left\lbrace \frac{1}{48\pi^2} \int_{4x_P}^1 \dd x (1-x)\left( 1 - \frac{4 x_P}{x}\right)^{3/2} \right. \\ \nonumber
    &  \times \left[\abs{\mathcal{F}_{PP}^{\tau\alpha}}^2 (1 + x - 2x^2) 
    + 6 \Re[\mathcal{F}_{PP}^{\tau\alpha}\,\mathcal{G}_{PP}^{\tau\alpha\ast}]\,(1-x) \right. \\ \nonumber
    &\left. \left. \quad + \abs{\mathcal{G}_{PP}^{\tau\alpha}}^2\left(\frac{2}{x} - x - 1\right)  \right]\right\rbrace\,,
\end{align}
where here $x = s/m_\tau^2$, with $s = (p_{P1} + p_{P2})^2$, the invariant mass of the two pseudoscalar system.

For the decays of pseudoscalar and vectors, in terms of $y_\alpha \equiv m_\alpha^2/m_P^2, \tilde{y} \equiv m_\alpha^2/m_V^2$
\begin{align}
    \label{eq:br_pseudoscalar}
    \mathrm{Br}(P \to \ell_\alpha \ell_\beta) &= \frac{\alpha_W^4}{64\pi}\frac{m_P^4}{M_W^4} \frac{m_P}{\Gamma_P} \frac{\abs{\mathcal{F}_P^{\beta\alpha}}^2}{m_P^2} y_\alpha \left(1 - y_\alpha\right)^2\,,\\[2ex]
    \label{eq:br_vector}
    \mathrm{Br}(V \to \ell_\alpha \ell_\beta) &= \frac{\alpha_W^4}{192\pi}\frac{m_V^4}{M_W^4} \frac{m_V}{\Gamma_V} \left[\frac{\abs{\mathcal{F}_V^{\beta\alpha}}^2}{m_V^4} \left(2 - \tilde{y}_\alpha - \tilde{y}_\alpha^2 \right) \right. \\ \nonumber
    & \quad + \frac{6\,\Re[\mathcal{F}_V^{\beta\alpha}\,\mathcal{G}_V^{\beta\alpha\ast}]}{m_V^4} \left(\tilde{y}_\alpha - \tilde{y}_\alpha^2 \right) + \left.\frac{\abs{\mathcal{G}_V^{\beta\alpha}}^2}{m_V^4}\left(\tilde{y}_\alpha + \tilde{y}_\alpha^2 - 2 \tilde{y}_\alpha^3 \right)\right] (1 - \tilde{y}_\alpha)\,,
\end{align}

For the definitions of $\mathcal{F}$ and $\mathcal{G}$, see Appendix~\ref{app:defintions_of_F_and_G}.

\subsection{Parametric suppression of meson decays}
The branching ratios of pseudoscalars and vectors decaying into two different charged leptons will be highly suppressed, because neutral unflavored mesons and quarkonia decay mostly through strong interactions. 
We can see this by analyzing the parametric dependence of Eqs.~\eqref{eq:br_pseudoscalar} and \eqref{eq:br_vector}
\begin{align}
    \mathrm{Br}(M \to \ell_\alpha \ell_\beta) \propto \left(\alpha_W^4\,\frac{m_M^4}{M_W^4} \frac{f_M^2}{m_M^2} \right)\frac{m_M}{\Gamma_M} \,.
\end{align}
The term in the parentheses will be much smaller than $m_M/\Gamma_M$, since $\Gamma_M \propto \alpha_s\,m_M$.\footnote{Neutral pion decay mostly to two photons through a chiral anomaly, so the parametric dependence of its total decay width should be $\Gamma_\pi \propto \alpha^2 m_\pi$. cLFV pion decays should therefore be less suppressed than, e.g., $\eta$ or $\eta^\prime$ decays. Nevertheless, they are still much smaller than $\tau$ decays.}

This is in direct contrast with $\tau$ decays, which will have a similar parametric dependence
\begin{align}
    \mathrm{Br}(\tau \to \ell_\alpha M) \propto \left(\alpha_W^4\,\frac{m_\tau^4}{M_W^4} \frac{f_M^2}{m_\tau^2} \right)\frac{m_\tau}{\Gamma_\tau}\,,
\end{align}
where the decay width scales as $\Gamma_\tau \propto \alpha_W^2\,m_\tau^5/M_W^4$. 
The small tau decay width cancels the small values of $\alpha_W^2\,m_\tau^4/M_W^4$.
 
Flavored mesons, like $K^0$, $D^0$, $B^0$, and $B_s^0$, decay mostly through weak interactions, and therefore these should not have the same suppression. 
However, the GIM mechanism should make their cLFV decays very small. 
In \cite{Awasthi:2024nvi}, some of these decays were investigated, and it was found that the decays are very suppressed, with $B$ and $B_s$ decays being larger than kaon decays.
This makes sense, since the large coupling between bottom and top quarks alleviates the GIM suppression. 

Heavier mesons, such as $\phi, J/\Psi$, and $\Upsilon$ have larger decay widths because they are subject to Okubo–Zweig–Iizuka suppression (OZI suppression) \cite{Okubo:1963fa, Zweig:1964jf, Iizuka:1966fk}. However, the decays are still mostly driven by strong interactions and are still parametrically suppressed.  

To finish our discussion about meson decays, it is also important to highlight the fact that pseudoscalar decays will also enjoy helicity suppression. 
Helicity suppression should be particularly relevant for heavy pseudoscalar states decaying to an $e\mu$ final state.

\subsection{Behavior of $\tau$ decays branching ratios with HNL parameters} \label{sec:behaviour_clfv}
Let us examine the behavior of different tau decays in terms of HNL parameters: $\Lambda_{\tau \beta}, U_{\mathrm{tot}}^2$, and $M_N$.

Considering as a first example the decay $\tau \to e \pi^0$, where we shall neglect the masses of both the pion and electron for now. 
Taking the formulas from Appendix~\ref{app:loop_formulas}, the branching ratio is
\begin{align}
    \mathrm{Br}(\tau \to e\pi^0) &\simeq 
    \frac{\alpha_W^4}{4096\pi} \frac{m_\tau^4}{M_W^4} \frac{m_\tau}{\Gamma_\tau} \frac{f_\pi^2}{m_\tau^2}\,\abs{2\,F_Z^{\tau e} + F_{\mathrm{u,box}}^{uu\tau e} + F_{\mathrm{d,box}}^{dd\tau e}}^2\,.
\end{align}

Let us examine the behavior of this function in two different limits: $M_N \ll M_W$ and $M_N \gg M_W$. 

\paragraph{Low mass limit $M_N \ll M_W$}
For HNLs much lighter than the EW scale, the branching ratio converges to (see the limits in Eqs.~\eqref{eq:loop_limits_1}--\eqref{eq:loop_limits_9})
\begin{equation}
    \mathrm{Br}(\tau \to e\pi^0) \xrightarrow{M_N \ll M_W}
    \frac{\alpha_W^4}{4096\pi} \frac{m_\tau^4}{M_W^4} \frac{m_\tau}{\Gamma_\tau} \frac{f_\pi^2}{m_\tau^2}\,\abs{\Lambda_{e \tau}}^2\,(U_{\mathrm{tot}}^2)^2  \frac{M_N^4}{M_W^4}\left[2 - \log(\frac{M_N^2}{M_W^2}) + \frac{1}{2}\,U_{\mathrm{tot}}^2 \right]^2\,,
\end{equation}
for small values of $M_N$, the branching ratio vanishes. 
This comes from a GIM suppression \cite{Lee:1977tib}, in the limit where $M_N \to 0$ the contributions coming from active neutrinos and HNLs cancel each other from the unitarity of $\mathcal{U}$.
The vanishing at low masses is universal among every cLFV process.

\paragraph{Large mass limit $M_N \gg M_W$}
For HNLs much heavier than the EW scale, $Z$ and $\gamma$ penguin diagrams dominate, and the contributions from box diagrams are negligible. In this limit
\begin{equation}
    \label{eq:high_mass_limit_tau_pion}
    \mathrm{Br}(\tau \to e\pi^0) \xrightarrow{M_N \gg M_W} \frac{\alpha_W^4}{4096\pi} \frac{m_\tau^4}{M_W^4} \frac{m_\tau}{\Gamma_\tau} \frac{f_\pi^2}{m_\tau^2} \abs{\Lambda_{e \tau}}^2\,(U_{\mathrm{tot}}^2)^2 \left[3 \log(\frac{M_N^2}{M_W^2}) - 5 + U_{\mathrm{tot}}^2\,\frac{M_N^2}{M_W^2} \right]^2\,. 
\end{equation}
Of particular importance to this expression is the $U_{\mathrm{tot}}^2\,M_N^2/M_W^2$ inside of the square brackets. 
This term implies that the branching ratio grows with mass; this behavior will be present in all branching ratios involving unflavored mesons, precisely due to the $Z$ penguin diagram. 
This \textit{non-decoupling} is also present in other cLFV decays not fully explored in this paper, such as $\ell_\alpha \to 3\ell_\beta$ or muon conversion in a nucleus (see e.g., \cite{Tommasini:1995ii, Urquia-Calderon:2022ufc})\footnote{Strictly speaking, every term here is non-decoupling since none of them vanish in the limit where $M_N\to\infty$. 
However, for the purposes of this paper, we shall use this inexact demarcation where we denote the non-decoupling part as part that depends on the second and above powers of $M_N$.}.

\begin{figure}[t]
    \centering
    \includegraphics[width=0.8\linewidth]{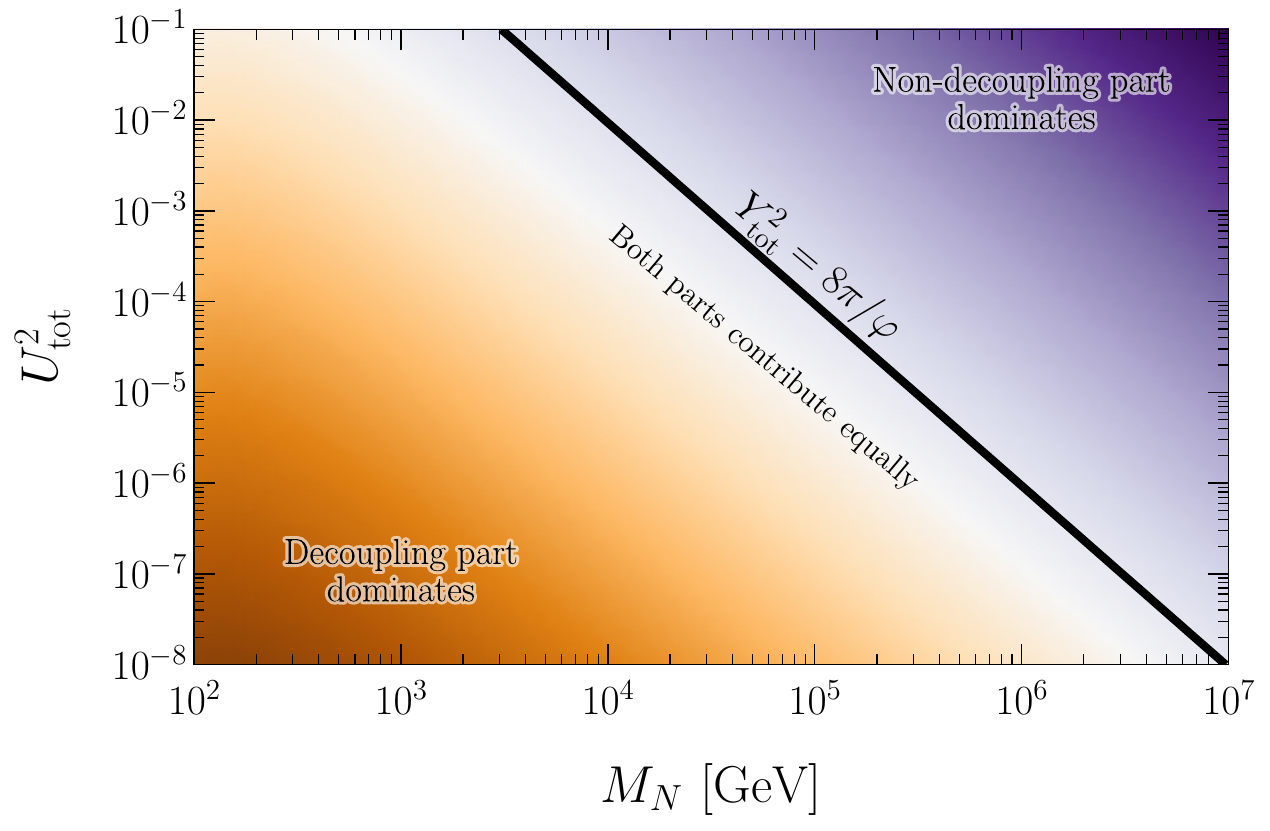}
    \caption{Regions of the parameter space where decoupling and non-decoupling. In the purplish regions is where the non-decoupling part dominates, the orange regions is where the decoupling part dominates, and in the white regions are where they are equally relevant. The black line shows the unitarity line from \cite{Urquia-Calderon:2024rzc}, where everything to the right of it is not allowed.}
    \label{fig:decoupling_v_non_decoupling}
\end{figure}

The non-decoupling behavior might seem unintuitive, as it goes against the Appelquist-Carazzone theorem \cite{Appelquist:1974tg}, that heavy particles inside loops should decouple. 
However, this is not the case for theories that under
go spontaneous symmetry breaking \cite{Veltman:1977kh, Toussaint:1978zm, Collins:1978wz} (see also Chapter 8 of \cite{Collins:1984xc}). 

In the SM, for example, we have non-decoupling because interactions with the Higgs and Goldstone bosons depend on the mass of the interacting particle. 
For example, in flavor-violating decays involving $b$ quarks, such as $b\to s\gamma$ or $B_S \to \mu\mu$ are relatively large because of the largeness of the $t$ quark \cite{Inami:1980fz, Hou:1986ug}.
Moreover, the $t$ quark mass does not decouple from its loop contributions in oblique parameters \cite{Peskin:1990zt, Peskin:1991sw}. 

In the minimal type-I seesaw, it happens for the same reasons. 
The interactions between neutral leptons and the scalar sector are proportional to the masses of fermions (see Eqs.~\eqref{eq:charged_goldstone_interactions}--\eqref{eq:higgs_interactions}). 
This happens because we can write the Yukawa matrices in terms of HNL masses and mixing angles.

We recover the decoupling behavior if we re-write Eq.~\eqref{eq:high_mass_limit_tau_pion} in terms of the total Yukawa coupling, $Y_{\mathrm{tot}}^2 = \frac{g^2 M_N^2}{2\,M_W^2}\,U_{\mathrm{tot}}^2$, instead of the total mixing angle \cite{DHoker:1984izu, DHoker:1984mif}. 
Indeed, the reason why this shape of non-decoupling does not spoil the use of effective field theory (EFT), whose backbone rests on the fact that heavy particles decouple, is that the non-decoupling only arises from the coupling.
Precisely in one-loop matching of the minimal type-I seesaw to SM operators, Yukawa couplings are ubiquitous \cite{Zhang:2021jdf}. 

In terms of the Yukawa, the branching ratio becomes
\begin{align}
    \mathrm{Br}(\tau \to e\pi^0) &\propto (Y_{\mathrm{tot}}^2)^2\,\frac{M_W^4}{M_N^4} \left[3 \log(\frac{M_N^2}{M_W^2}) - 5 + \frac{2}{g^2}\,Y_{\mathrm{tot}}^2 \right]^2 + \mathcal{O}\left(\frac{M_W^6}{M_N^6} \right)\,.
\end{align}
For a fixed Yukawa coupling, and for $M_N \gg M_W$, all branching ratios vanish, in accordance with the decoupling theorem. 
The Yukawa coupling cannot be arbitrarily large since it would compromise both perturbativity and tree-level unitarity. 
To make sure that tree-level unitarity is satisfied, we will restrict the Yukawa coupling to $Y_{\mathrm{tot}}^2 < 8\pi/\varphi$, where $\varphi$ is the golden ratio \cite{Urquia-Calderon:2024rzc}.

The \textit{non-decoupling} and \textit{decoupling} both become relevant when
\begin{align}
    U_{\mathrm{tot}}^2 \frac{M_N^2}{M_W^2} \simeq  \log(\frac{M_N^2}{M_W^2})\,,
\end{align}
for larger values of $U_{\mathrm{tot}}^2$, the non-decoupling part dominates; but for smaller values the decoupling part dominates. 
The non-decoupling part will never dominate in cLFV processes without first compromising tree-level unitarity. 
This is universal for all cLFV processes with non-decoupling. 
See Fig.~\ref{fig:decoupling_v_non_decoupling} to see in which regions of the $M_N, U_{\mathrm{tot}}^2$ plot the non-decoupling portion is relevant.

We plot the behavior of different branching ratios with respect to $M_N$ for a fixed $U_{\mathrm{tot}}^2$. 
See Fig.~\ref{fig:tau_decay_predictions_IO}.

\section{Predictions and indirect bounds}
\label{sec:predictions}
\begin{figure}[t]
    \centering
    \includegraphics[width = \linewidth]{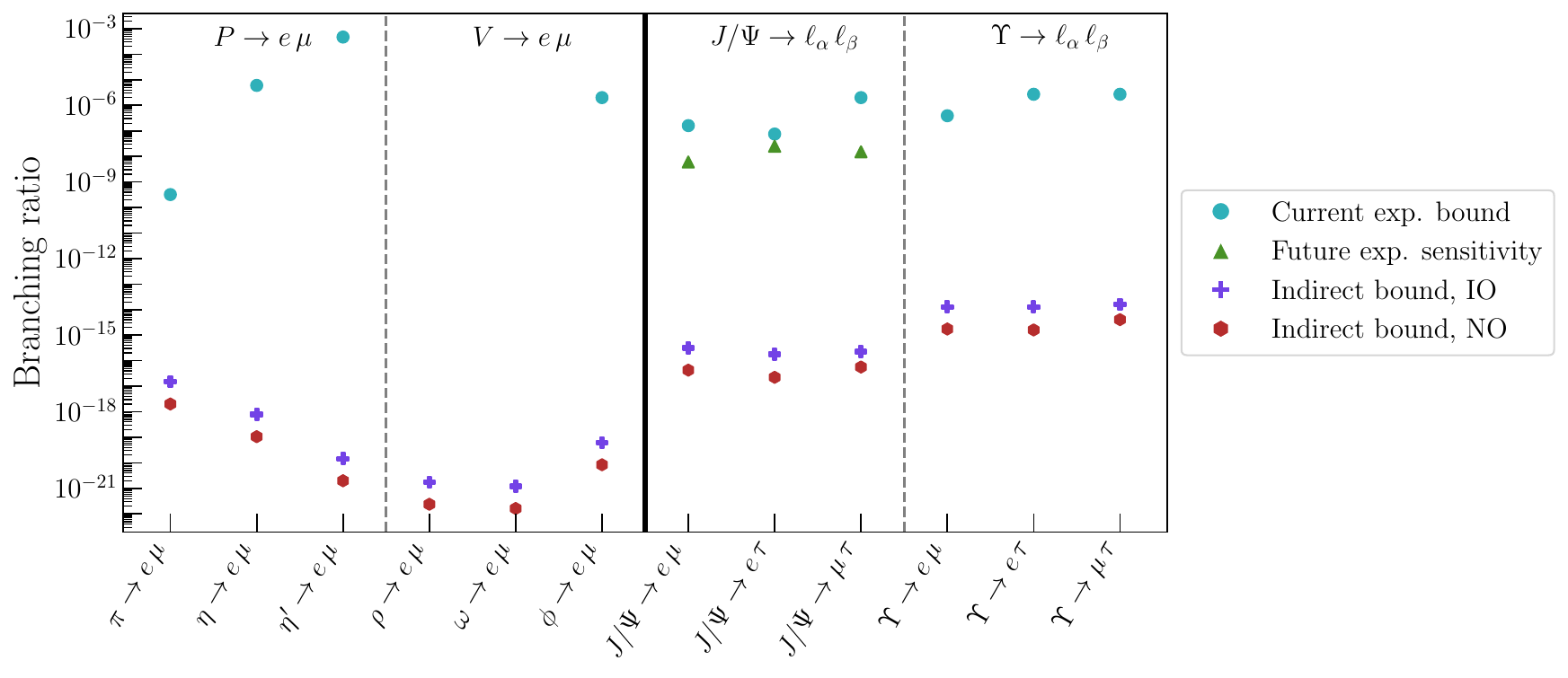}
    \includegraphics[width = \linewidth]{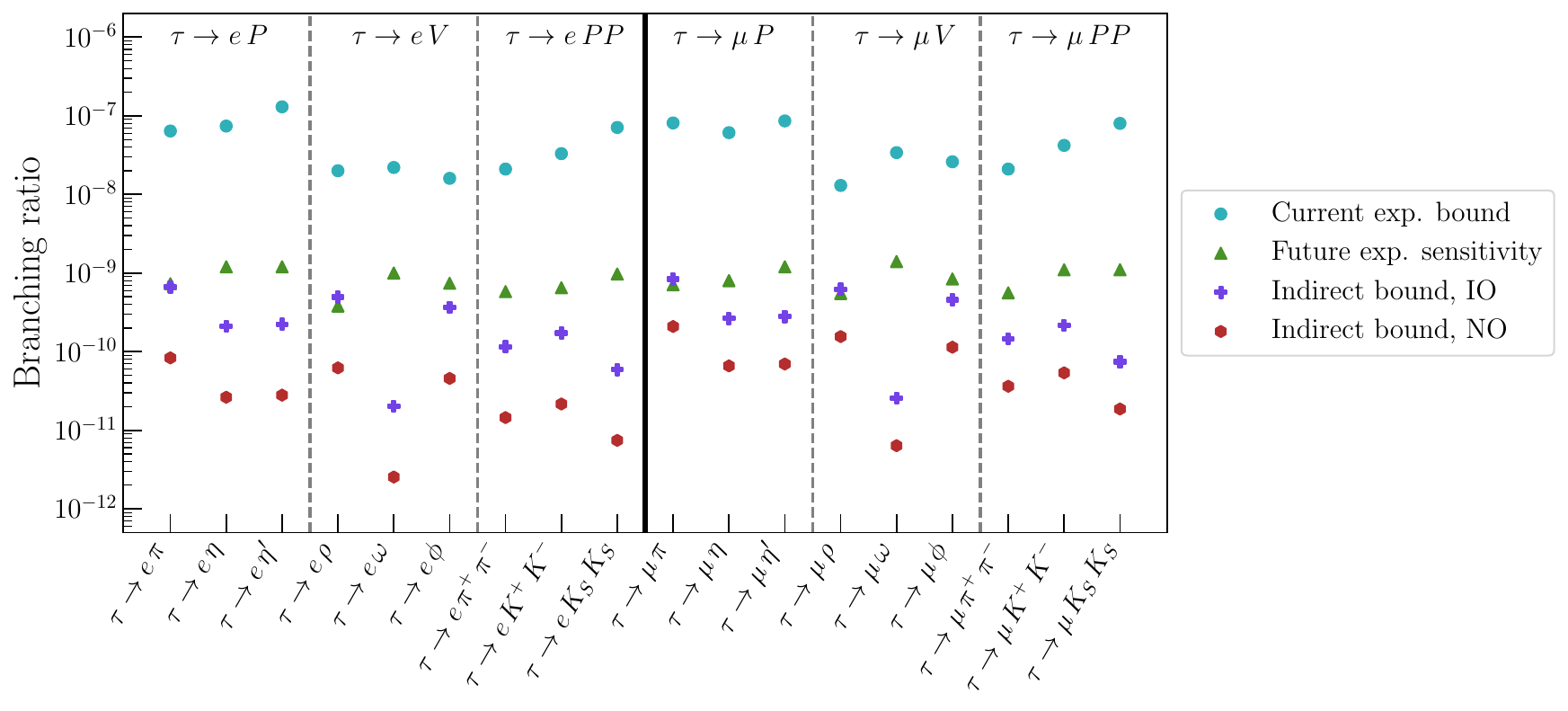}
    \caption{Comparison between the maximum allowed branching ratios, and current and prospective searches. For the values of current and prospective searches, see Table~\ref{tab:cLFV_decays_limits}.
    We used the values that maximize each branching ratio (see main text).}
    \label{fig:decay_bound_comparison}
\end{figure}

We determine indirect upper bounds on different cLFV processes, where we used the HNL parameters that maximize each individual branching ratio. 
Processes in the $e\mu, e\tau$, and $\mu\tau$ sectors are maximal when $\abs{\Lambda_{e\mu}}, \abs{\Lambda_{e\tau}}$ and $\abs{\Lambda_{\mu\tau}}$ are the largest. The maximal allowed values are restricted by neutrino oscillation data and are summarized in Table~\ref{tab:allowed_ratios}.

The \textit{maximal} allowed value of $U_\mathrm{tot}^2$ is determined by experimental and indirect constraints.
For $M_N$ larger than the electroweak scale, only indirect bounds matter.
These indirect bounds are driven by non-unitarity of the PMNS matrix, lepton universality constraints, and other cLFV processes \cite{Blennow:2023mqx}.
For NO $U_{\mathrm{tot}}^2 < \num{5.8e-4}$, and for IO $U_{\mathrm{tot}}^2 < \num{1.2e-3}$.

As we explained in the previous section, the non-decoupling makes the branching ratios grow with $M_N$ at fixed $U_{\mathrm{tot}}^2$. 
We will choose the maximal value of $M_N$ allowed by tree-level unitarity \cite{Urquia-Calderon:2024rzc} and by indirect constraints, which should be around
\begin{equation}
    \begin{aligned}
        U_{\mathrm{tot}}^2\,M_N^2 &\leq \left(\SI{686.28}{GeV}\right)^2\,, \\
        \implies M_N^{\mathrm{max}} &= \left\lbrace \begin{aligned}
            \SI{28.50}{TeV}&&&\text{for NO}\,, \\
            \SI{19.81}{TeV}&&&\text{for IO}\,.
        \end{aligned}\right.
    \end{aligned}
\end{equation}
This bound is valid as long as we have large mixing angles, above the seesaw line $U_{\mathrm{tot}}^2 \simeq m_\nu/M_N$, where the bound saturates.

Figure~\ref{fig:decay_bound_comparison} shows the maximal values of each branching ratio, and compares them with current experimental bounds and prospective sensitivities. 

As we expected, meson decays are very suppressed, and our indirect bounds are much more stringent than prospective searches, some of which are around the same order of magnitude as the processes explored in \cite{Awasthi:2024nvi}. 
Heavier quarkonia ($J/\Psi, \Upsilon$) decays have larger allowed branching ratios from the OZI rule, and particularly $\Upsilon$ decay will have an enhancement from top non-decoupling. 
Nonetheless, the experimental bounds and the future sensitivity of BES-III remain much larger than our theoretical bounds.

\begin{figure}
    \centering
    \includegraphics[width = \linewidth]{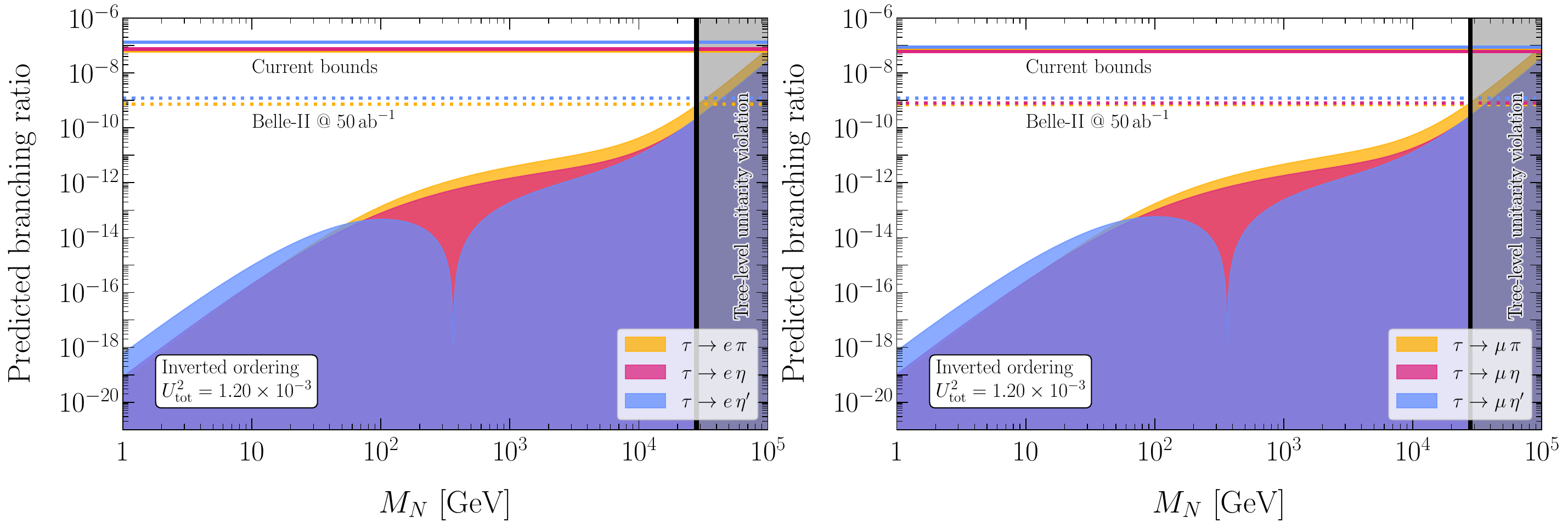}
    \includegraphics[width = \linewidth]{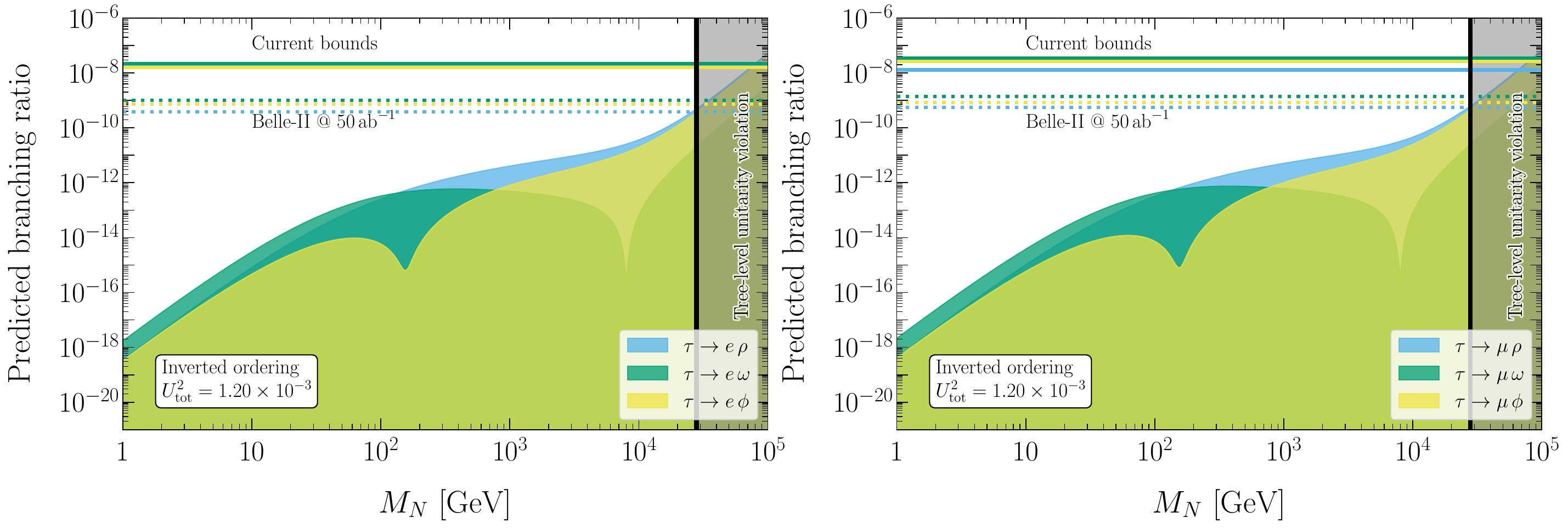}
    \includegraphics[width = \linewidth]{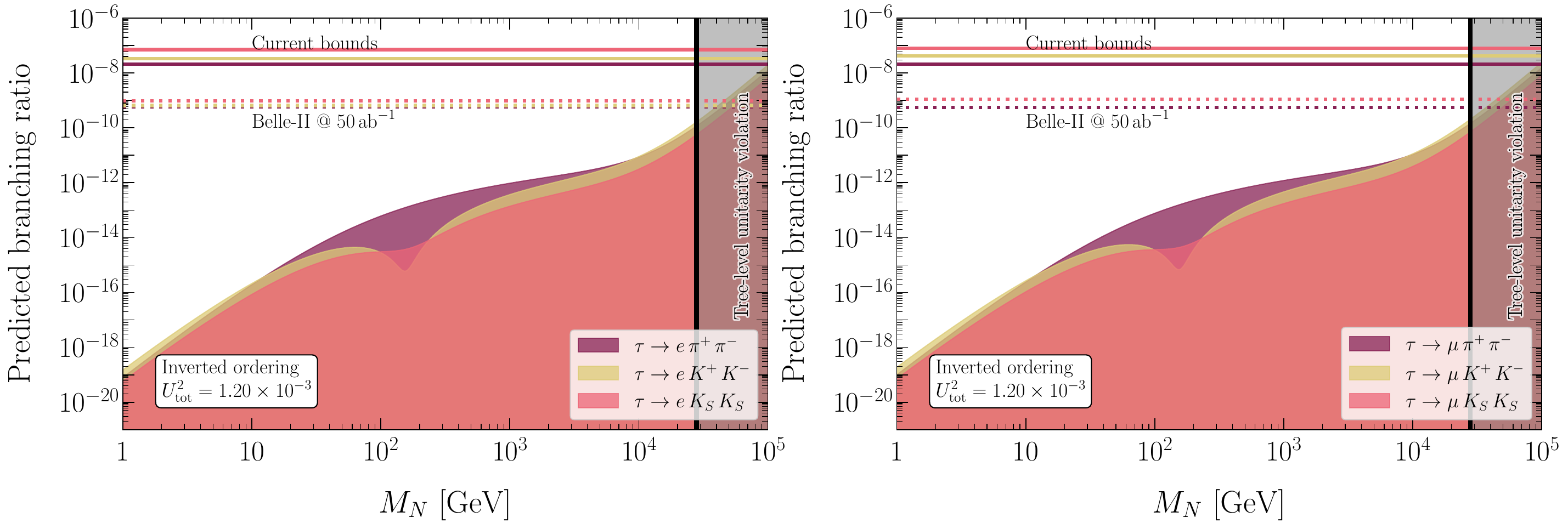}
    \caption{Maximal branching ratio for different $\tau$ decays in IO, compared with current bounds and prospective searches.}
    \label{fig:tau_decay_predictions_IO}
\end{figure}

$\tau$ decays have much larger allowed branching ratios. 
In particular, our theoretical upper bounds of $\tau \to \ell_\alpha\,\rho$ and $\tau \to \ell_\alpha\,\pi$ are in the ballpark of the expected sensitivity of Belle-II. 

An observation by Belle-II of either process could be compatible with the minimal type-I seesaw. 
We must stress this would be in a very fortunate region of the parameter space, where we have IO, and $U_{\mathrm{tot}}^2, M_N$, and $\abs{\Lambda_{\tau\alpha}}$ are conspiring to enhance either process.
In the case of observation (and also in the case that indirect bounds do not get stronger by the time Belle-II finishes collecting data, which will probably not be the case), a much more delicate analysis should be made to verify if this model would be a source for it.

In Fig.~\ref{fig:tau_decay_predictions_IO} we show the maximal allowed branching ratio with fixed $U_{\mathrm{tot}}^2$ with varying $M_N$ and varying $\abs{\Lambda_{\alpha\beta}}$.
Since in IO, all of them are allowed to be zero, the bands go all the way down to zero. For NO, we will see finite bands.
Fig.~\ref{fig:tau_decay_predictions_IO} shows a better picture of the allowed branching ratio for different values of $M_N$.
For smaller values of $M_N$, we would be expecting much smaller allowed values of the branching ratio, far from the prospective sensitivity of any experiment.

It would also be interesting to perform the same exercise within other seesaw models. 
For example, in the type-I seesaw with three degenerate HNLs, the global bounds from indirect measurements are less constraining and will therefore provide less constraining indirect bounds on the branching ratios, therefore making it more likely for it to be seen in Belle-II.

\section{Bounds from $\tau$ decays and comparison between decays} \label{sec:bounds}
\begin{figure}[t]
    \centering
    \includegraphics[width = \linewidth]{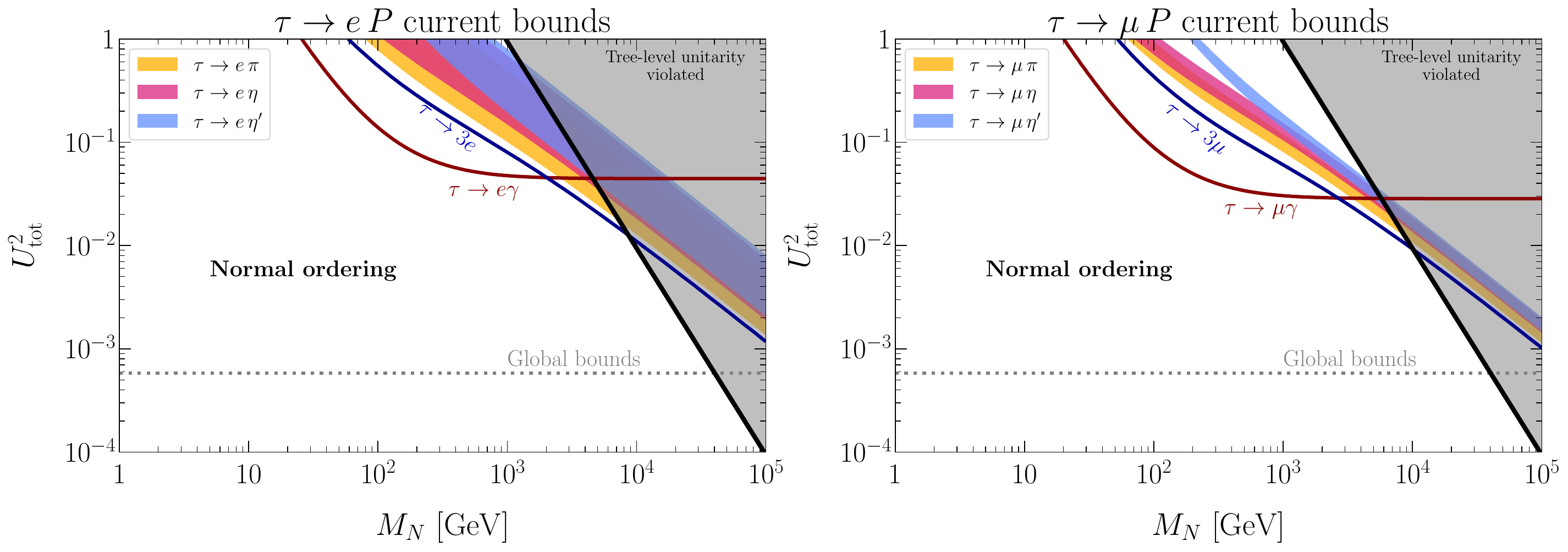}
    \includegraphics[width = \linewidth]{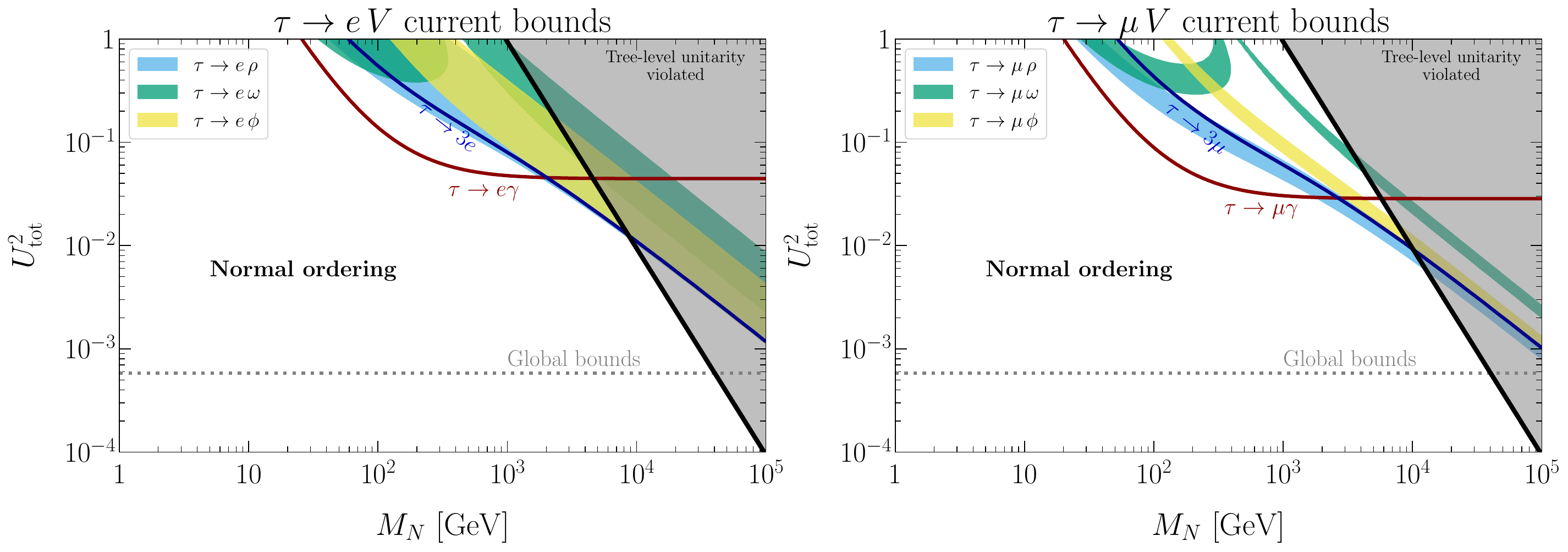}
    \includegraphics[width = \linewidth]{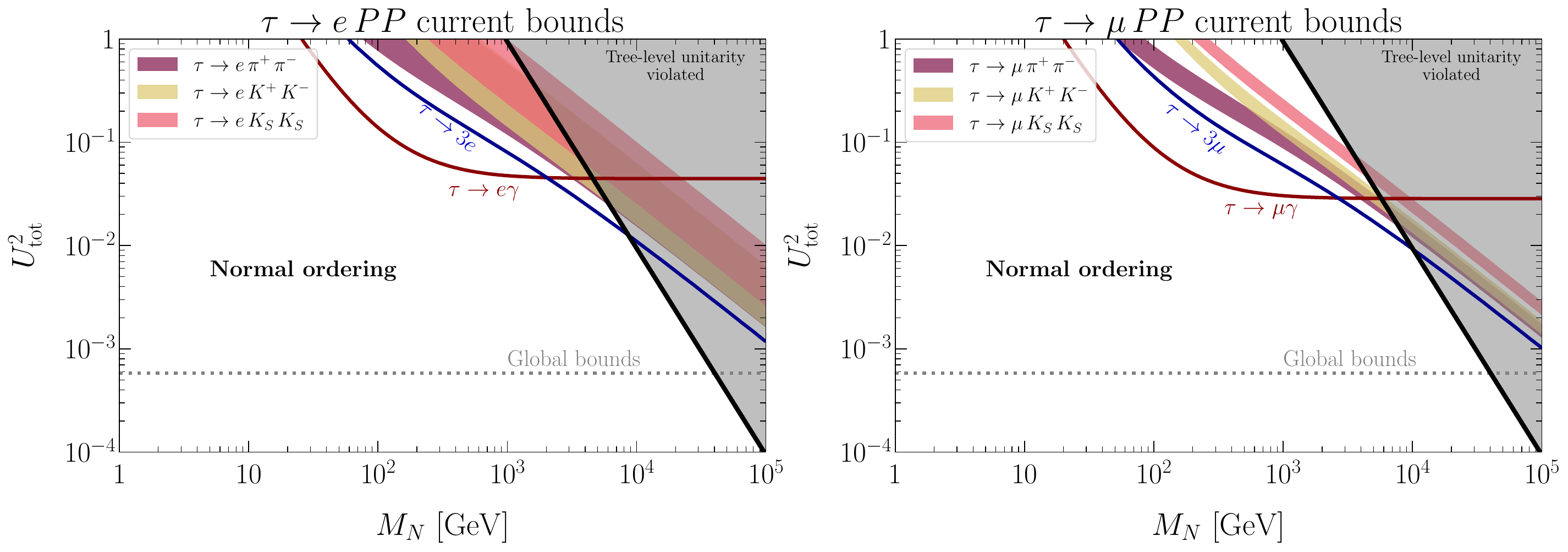}
    \caption{Bounds on HNL parameters from different cLFV $\tau$ decays to different meson systems. 
    The band represents anywhere where the branching ratio is equal to its current experimental bound, listed in Table~\ref{tab:cLFV_decays_limits}, and is allowed by neutrino oscillation data (see Fig.~\ref{fig:flavor_triangle}, or Table~\ref{tab:allowed_ratios}). 
    \textbf{The $\tau \to \ell\gamma$ and $\tau\to 3\ell$ bounds do not show a band, but only show the line where the bounds would be maximized, and should be compared only with the lowest part of the band of each bound.}}
    \label{fig:normal_ordering_bounds}
\end{figure}
\begin{figure}[t]
    \centering
    \includegraphics[width = \linewidth]{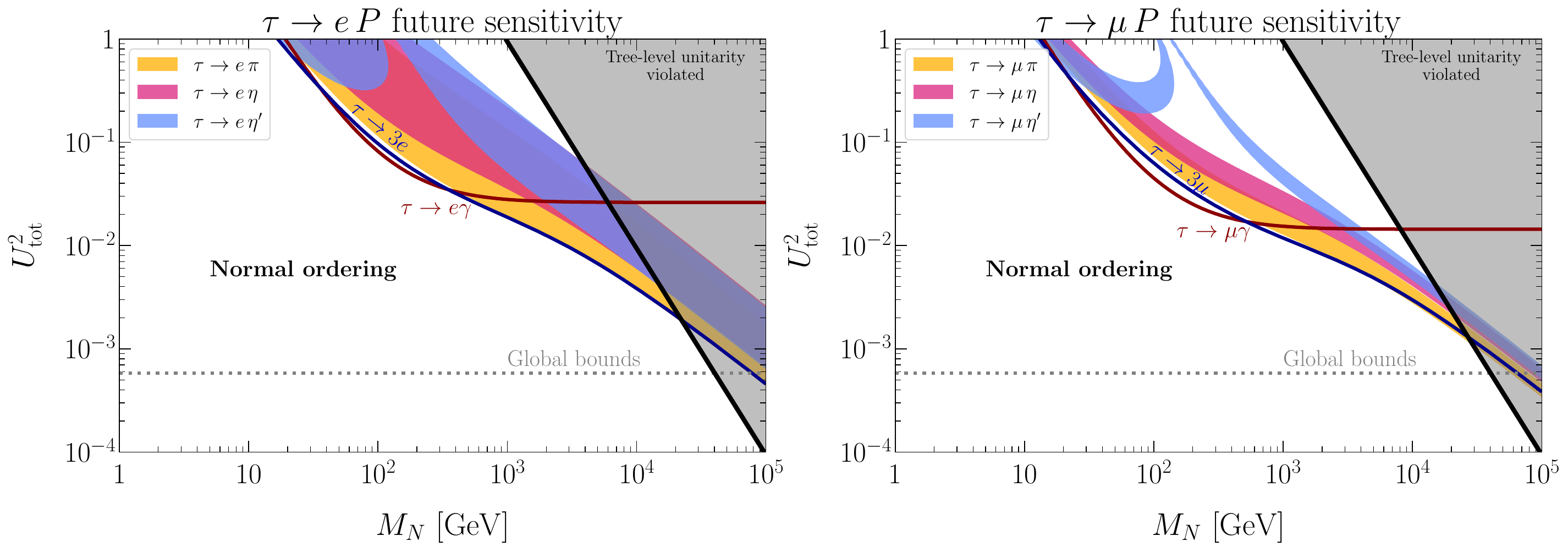}
    \includegraphics[width = \linewidth]{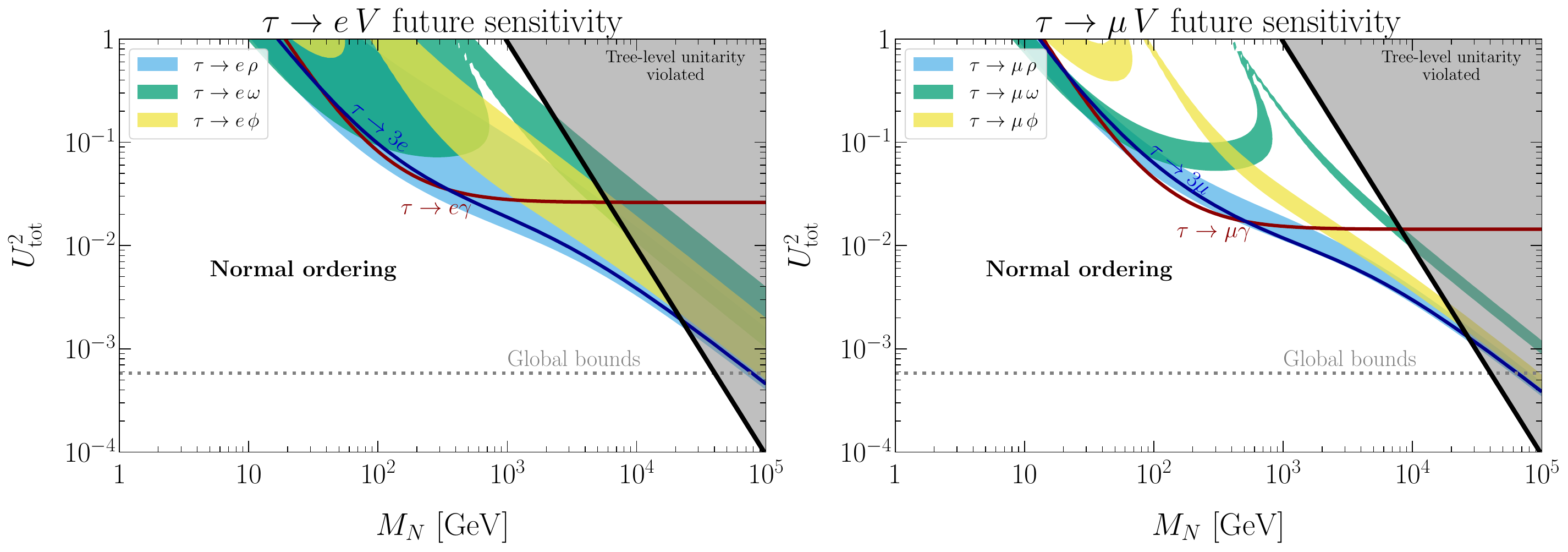}
    \includegraphics[width = \linewidth]{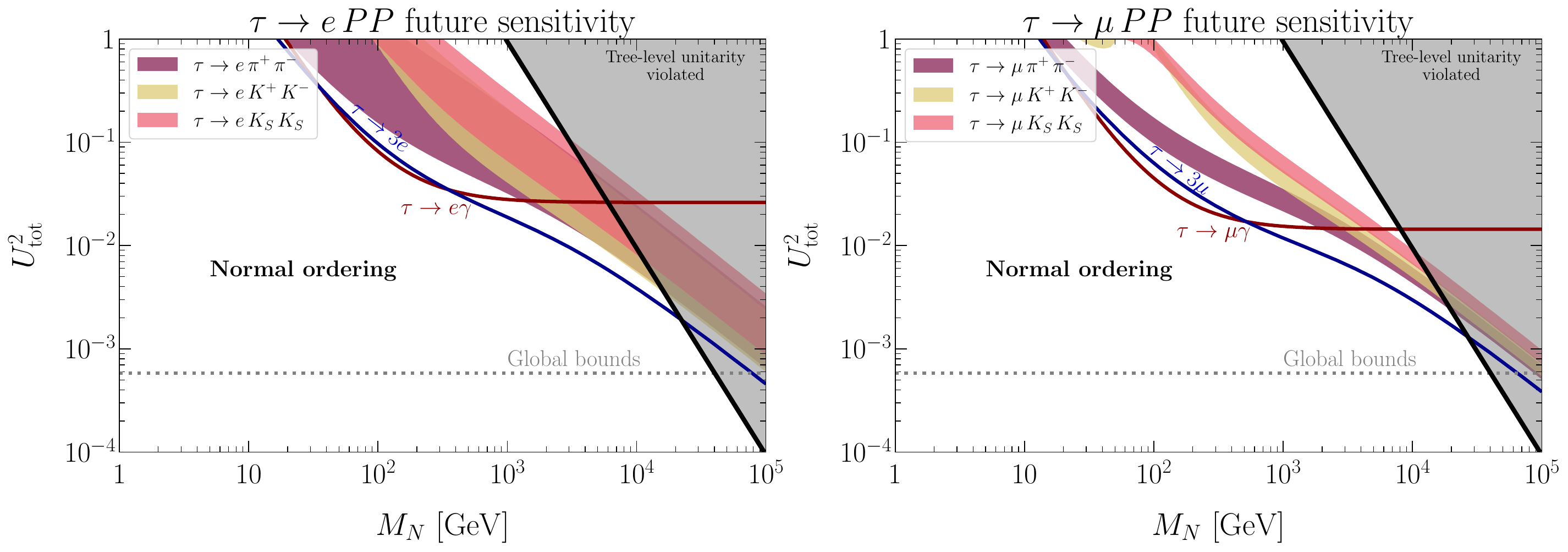}
    \caption{Future sensitivity from proposed experiments. See the caption of Fig.~\ref{fig:normal_ordering_bounds} and the main text for details.}
    \label{fig:normal_ordering_future}
\end{figure}

We will now provide the current bounds that the non-observation of different processes provides, and compare them with other processes in the $\tau$ sector. 

We show the bounds in Figure~\ref{fig:normal_ordering_bounds} for NO the plots for IO are in Appendix~\ref{app:additional_plots} in Fig.~\ref{fig:inverted_ordering_bounds}.

The bands shown represent the values of parameter space where the branching ratio equals its current bound for different values of $\Lambda_{\tau\alpha}$ as allowed by neutrino oscillation data. This represents that the bounds are weaker for smaller values $\Lambda_{\tau_\alpha}$.
The lower part of each band is where each $\abs{\Lambda_{\alpha\tau}}$ is maximum, and therefore provides the most optimistic bounds, and the upper part represents where $\abs{\Lambda_{\alpha\tau}}$ is minimum.

We also compare these bounds to other $\tau$ cLFV decays that are more studied in the literature, $\tau \to \ell_\alpha \gamma$, and $\tau \to 3\ell_\alpha$. 
The formulas for these processes are widely available in the literature; therefore, we shall not repeat them (see e.g. \cite{Ilakovac:1994kj}). 
For these decays, we use the most recent bounds found in the literature, listed also in Table~\ref{tab:cLFV_decays_limits}. 
In Figure~\ref{fig:normal_ordering_bounds}, for these two processes, we show the line where $\abs{\Lambda_{\tau\alpha}}$ is maximum, and \textbf{therefore only the lower part of each band can be compared with the $\tau \to \ell \gamma$ and $\tau \to 3\ell$ lines.}

We also show the prospective sensitivity of future searches for NO in
Fig.~\ref{fig:normal_ordering_future} (and for IO in Fig.~\ref{fig:inverted_ordering_future}).

The previous section provided hints that the largest branching ratio comes from $\tau \to \ell_\alpha\,\pi$ and $\tau \to \ell_\alpha\,\rho$ decays. 
We can appreciate it again in  Figs.~\ref{fig:normal_ordering_bounds} and \ref{fig:normal_ordering_future}, where these $\tau$ decays can potentially dominate the future searches in the tau sector.

\begin{figure}[t]
    \centering
    \includegraphics[width=0.45\linewidth]{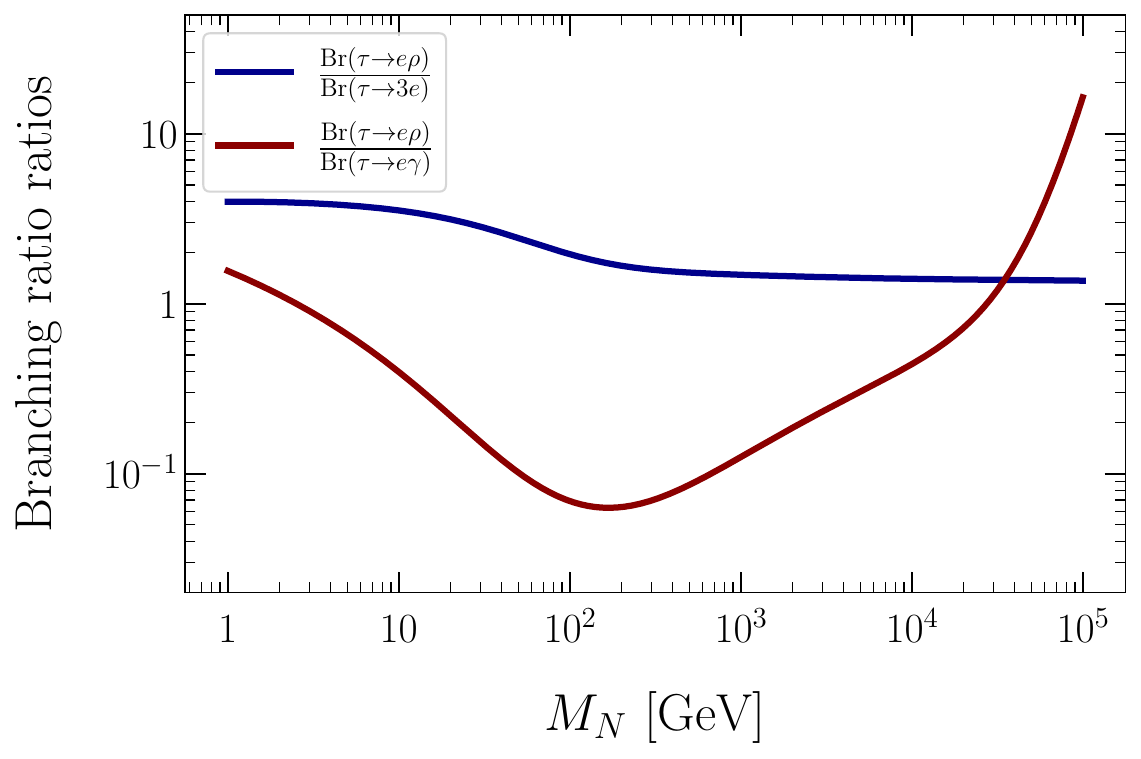}%
    \includegraphics[width=0.45\linewidth]{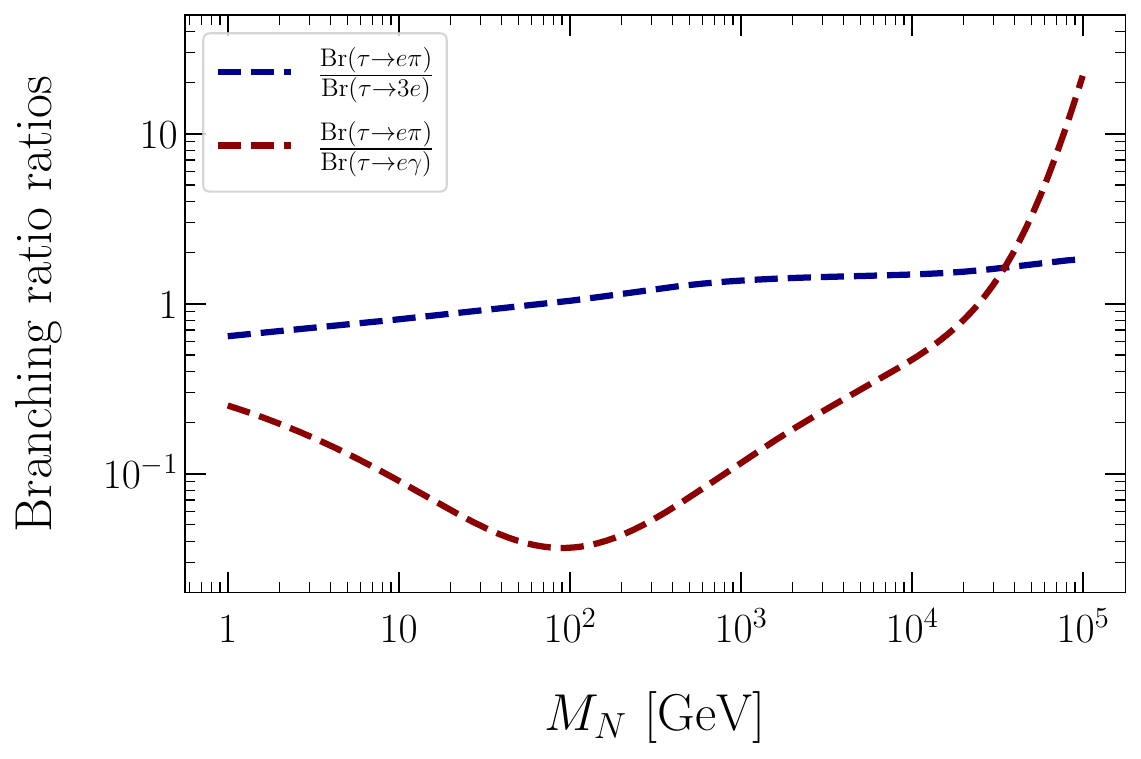}
    \caption{Ratio between $\tau \to e\rho$ (left) and $\tau \to e\pi$ (right) with the canonical $\tau$ decays.}
    \label{fig:ratios_rho_pi}
\end{figure}

To explain how these processes can compete, we can make a naive estimate the ratio, $\frac{\mathrm{Br}(\tau \to \ell \rho)}{\mathrm{Br}(\tau \to 3\ell)} \propto 16\pi^2\,\left(\frac{f_\rho}{m_\tau}\right)^2 \simeq 2.26$. 
Indeed, despite the relative suppression from the energy scale of the form factor compared to the $\tau$, this is counteracted by the phase space suppression of the three-body decay. 
A more careful calculation reveals that this ratio is closer to 4, for lighter values of $M_N$, and it converges close to 1.5 for larger $M_N$, as shown in Fig.~\ref{fig:ratios_rho_pi}. 

For decay to pions, on the other hand $\frac{\mathrm{Br}(\tau \to \ell \rho)}{\mathrm{Br}(\tau \to 3\ell)} \propto 16\pi^2\,\left(\frac{f_\pi}{m_\tau}\right)^2 \simeq 0.8$, which implies that both decays should be around the same order of magnitude (we are repeating this statement from \cite{deGouvea:2013zba}).
We show how this ratio changes for different values of $M_N$ in the right plot in Fig.~\ref{fig:ratios_rho_pi}.\footnote{A similar result was obtained in the context of superymmetric theories, where $\tau\to \mu\eta$ is predicted to be larger than $\tau \to 3\mu$ and $\tau \to \mu\gamma$.}

The largeness of $\tau \to e + \text{meson}$ compared to $\tau \to e\gamma$ should not come as a surprise, since the $\tau \to e\gamma$ does not decouple for heavy HNLs, and it plateaus logarithmically.
This explains the dominance of semi-leptonic $\tau$ decays at large masses.

\subsection{Comparison between cLFV muon decays, and $\tau\to \ell \pi$ $\tau\to \ell \rho$}
\begin{figure}
    \centering
    \includegraphics[width=\linewidth]{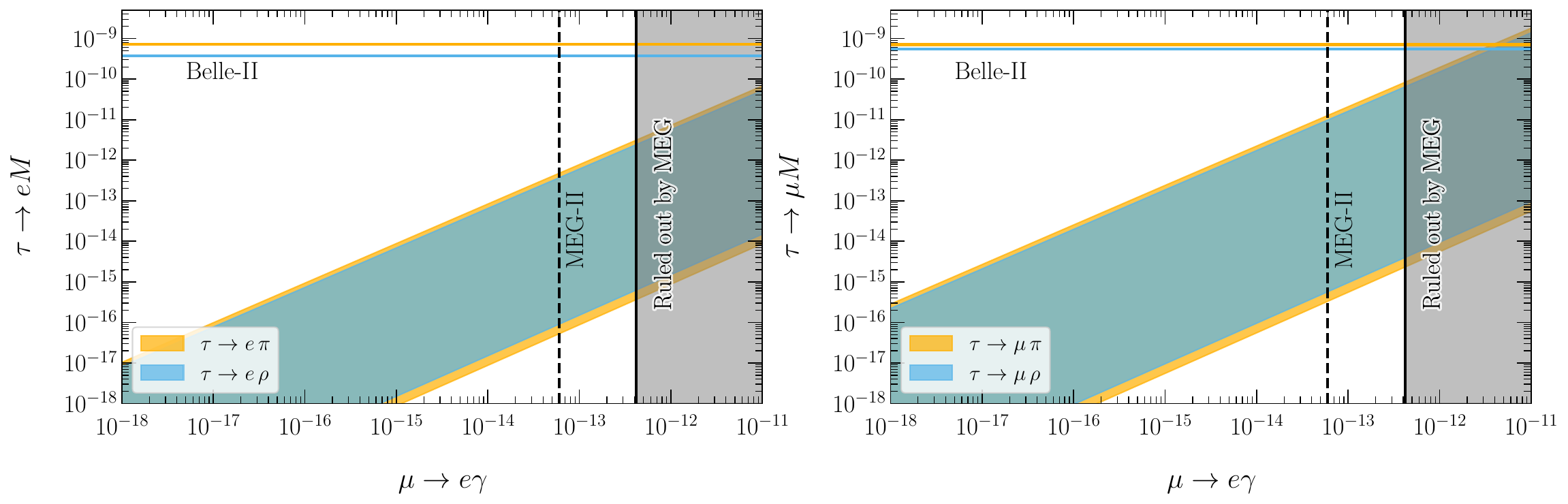}
    \includegraphics[width=\linewidth]{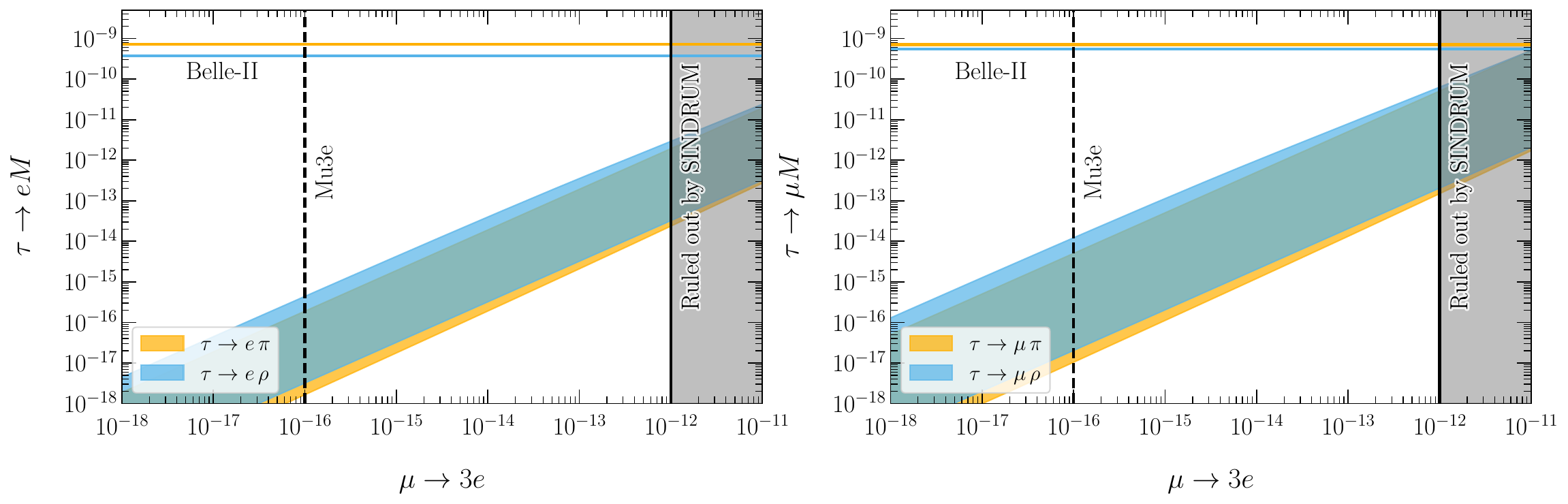}
    \caption{Upper two plots: allowed values of $\tau \to \ell \pi$ and $\tau\to\ell\rho$ (with $\ell = e$ in the right, and $\ell = \mu$ in the left) as a function of $\mu\to3e$, when marginalized over HNL parameters with $m_N \in \left[1, 10^7\right]\si{GeV}$, $U_{\mathrm{tot}}^2 \in \left[10^{-7}, 10^{-2} \right]$, over PMNS parameters in the normal ordering (see Fig.~\ref{fig:flavor_triangle} and Table~\ref{tab:allowed_ratios}), and as long as perturbativity holds. 
    We also showed ruled out regions of parameter space by MEG, and the sensitivities of MEG-II and Belle-II.
    Lower two plots: Same as two upper plots but with $\mu\to 3e$, and their respective sensitivities.}
    \label{fig:muon_decays_vs_tau_decays}
\end{figure}
Two of the strongest constraints of the model come from bounds of the non-observation of $\mu\to e\gamma$ and $\mu\to 3e$. 
The analytical ratios between these muon cLFV decays and other $\tau$ decays, namely $\tau \to \ell\gamma$ and $\tau\to 3\ell$, were previously obtained in the literature \cite{Gavela:2009cd, Dinh:2013vya}.
While similar analytic relations can be derived within our framework, the semileptonic $\tau$ decays considered here receive contributions from different linear combinations of loop functions, making compact expressions less transparent. 
For this reason, we adopt a numerical approach that fully captures the underlying flavor correlations.

The MEG collaboration has placed the strongest bounds on $\mathrm{Br}(\mu\to e\gamma) < \num{4.2e-13}$ \cite{MEG:2016leq}, and the MEG-II experiment will aim to probe branching ratios down to $\num{6e-14}$\cite{MEGII:2018kmf}.
On the other hand, searches of by SINDRUM have placed bounds of $\mathrm{Br}(\mu\to 3e
) < \num{1.0e-12}$ \cite{SINDRUM:1987nra}, and the Mu3e collaboration will aim to reach branching ratios of $\num{1e-16}$ \cite{Berger:2014vba}. 
Following our discussion from Sec.~\ref{sec:behaviour_clfv}, the branching ratio of $\mu\to e\gamma$ at one loop will plateau logarithmically with HNL mass, but $\mu\to 3e$ will grow with HNL mass, at a fixed mixing angle. 
The analytical formulas for these branching ratios can be found in, e.g.,~\cite{Urquia-Calderon:2022ufc}.

In a Sec.~\ref{sec:predictions} we derived a \textit{theoretical} upper bound on cLFV semileptonic $\tau$ decays and meson decays, by using the maximum total mixing angle as allowed by global bounds~\cite{Blennow:2023mqx}, the maximal mass as allowed by perturbativity considerations~\cite{Urquia-Calderon:2024rzc}, and the maximum flavor mixing as allowed by neutrino oscillation data (see Fig.~\ref{fig:flavor_triangle} and Table~\ref{tab:allowed_ratios}).
We will once again perform a similar exercise, but using the strong bounds from muon decays. 
It is very important to stress that we cannot do this in the inverted ordering because the model allows $\mu\to e\gamma$ to exactly vanish, as $\Lambda_{e\mu} = 0$ from oscillation data \cite{Dinh:2013vya}\footnote{This is as long as we are in the limit where $M_{N_1} = M_{N_2}$ and $\Theta_{\alpha 1} = i\Theta_{\alpha 2}$. Departure from these equalities can generate a non-zero $\Lambda_{e\mu}$ \cite{Ruchayskiy:2011aa}.}, 
and therefore would not allow us extract any meaningful bounds.

We show in Fig.~\ref{fig:muon_decays_vs_tau_decays} the allowed values of the processes with the largest branching ratios we explored in this paper: $\tau\to \ell\pi$ and $\tau\to\ell\rho$. 
We obtained this region by sampling points for different HNL masses $m_N \in \left[1, 10^7\right]\,\si{GeV}$, over different angles $U_{\mathrm{tot}}^2 \in \left[10^{-7}, 10^{-2}\right]$, $\Lambda_{\alpha\beta}$ as allowed by oscillation data, and restricting parameters to respect tree-level perturbativity. 

The current bounds from $\mu\to e\gamma$ and $\mu\to 3e$ restrict the branching ratios of $\tau\to e\pi$ ($\tau\to \mu\pi$) to be less than \num{2.346e-12} (\num{5.076e-11}) and the ones of $\tau\to e\rho$ ($\tau\to\mu\rho$) to be less than \num{1.820e-12} (\num{4.896e-11}), providing stronger bounds than the ones we obtained considering global bounds.

All these branching ratios are below the expected sensitivity of Belle-II.
Mu3e experiment, in case of discovery of $\mu\to 3e$, will be capable of  providing us a band seen in Fig.~\ref{fig:muon_decays_vs_tau_decays} for a prediction for the rare tau decays we explored in this paper. In the case of non-discovery, it will provide the strongest \textit{theoretical} constraints on the model.

\section{Summary and conclusions} \label{sec:conclusions}
We studied different charged lepton flavor violating processes involving mesons in the minimal type-I seesaw with two degenerate HNLs.

We find that meson decays, including light-unflavored mesons and quarkonia, have very small branching ratios in our model.
The largest of light-unflavored mesons are pion decays, where the branching ratio is at least $\num{e-17}$ as allowed by indirect and theoretical constraints; 
and for quarkonia $\Upsilon$ decays are as large as $\num{e-14}$. We also explained the reason for their suppression.

$\tau$ decays into mesons, on the other hand, are much larger. 
In particular, $\tau \to \pi\,\ell_\alpha$ and $\tau \to \rho\,\ell_\alpha$ have the largest allowed branching ratios, in the order of $\num{e-10}$. 
This is in the ballpark of the sensitivity that Belle-II aims to achieve. 

However, it should still be unlikely that Belle-II will see anything, since it is likely that electroweak precision bounds will improve in the coming years with upcoming experiments like PIONEER \cite{PIONEER:2022yag}, which will provide more stringent bounds on electroweak precision bounds from lepton universality decays on pions.

The currently running and upcoming muon facilities \cite{Mu2e:2014fns, Berger:2014vba, COMET:2018auw, MEGII:2018kmf} will also provide the strongest bounds ever seen in any cLFV process. 
In the case of observation of any cLFV process in these current and upcoming facilities, we should then aim to reconstruct mixing angles and masses to pinpoint which regions of the parameter space we could lie at.

It would also be interesting to perform the same exercise with other seesaw variants, like the minimal type-I seesaw with three degenerate HNLs, seesaw models with two or more additional scales, like the inverse or double seesaw.
The bounds from these models are less constraining, and should therefore open up more the possibility of seeing cLFV processes with mesons future facilities, especially Belle-II.

\acknowledgments
We are grateful to Markus Ahlers, Stefan Antusch, and Jacobo López-Pavón for their comments on KAUC's PhD thesis, to Xabier Marcano for his thorough review of an early manuscript, and to Juan Carlos Helo for his comments on early manuscript. 
We are also indebted to Merlijn van Emmerick for helpful discussions and sharing her Master's thesis results with us, which covered a subset of the processes examined in this paper.

%This project has received funding from the European Research Council (ERC) under the European Union’s Horizon 2020 research and innovation programme (Program No. GA 694896) and from the Carlsberg Foundation (grant agreement CF17-0763).

\appendix
\section{Unitarity and seesaw relations}
\label{app:unitarity_seesaw_formulas}
The $\mathcal{U}$ and $\mathcal{C}$ matrices introduced in Section~\ref{sec:theory} follow a set of equalities that are a consequence of the seesaw relation (see Eq.~\eqref{eq:seesaw}), and from the unitarity of the matrix that diagonalizes the mass matrix in Eq.~\eqref{eq:mass_lagrangian}. For $\mathcal{N}$ HNLs, the relations are
\begin{align}
    \label{eq:definition_C}
    \sum_{\alpha = e,\mu,\tau}\mathcal{U}_{\alpha i}^{\ast}\,\mathcal{U}_{\alpha j}^{\phantom{\ast}} &= \mathcal{C}_{ij}\,, \\
    \label{eq:unitarity_of_U}
    \sum_{i = 1}^{\mathcal{N} + 3}\mathcal{U}_{\alpha i}^{\phantom{\ast}}\,\mathcal{U}_{\beta i}^\ast &= \delta_{\alpha\beta}\,, \\
    \label{eq:more_equations}
    \sum_{i = 1}^{3 + \mathcal{N}} \mathcal{U}_{\alpha i}^{\phantom{\ast}}\,\mathcal{C}_{ij}^{\phantom{\ast}} &= \mathcal{U}_{\alpha i}\,,\hspace{2cm}
    \sum_{i = 1}^{3 + \mathcal{N}} \mathcal{C}_{ij}^{\phantom{\ast}}\,\mathcal{C}_{ik}^{\phantom{\ast}} = \mathcal{C}_{kj}\,,& \\
    \label{eq:seesaw_relations_reloaded}
    \sum_{i = 1}^{3 + \mathcal{N}} \mathcal{U}_{\alpha i}^{\phantom{\ast}}\,\mathcal{U}_{\beta i}^{\phantom{\ast}}\,m_i &= \sum_{i = 1}^{3 + \mathcal{N}} \mathcal{U}_{\alpha i}^{\phantom{\ast}}\,\mathcal{C}_{ij}^{\ast}\,m_i = 
    \sum_{i = 1}^{3 + \mathcal{N}} \mathcal{C}_{ij}^{\phantom{\ast}}\,\mathcal{C}_{ik}^{\phantom{\ast}}\,m_i = 0\,,
\end{align}
where here the $i$ index goes through all active neutrinos and $\mathcal{N}$ HNLs, including the $m_i$, which can represent either the mass of neutrinos or HNLs. 
Eq.~\eqref{eq:definition_C} is the definition of the $\mathcal{C}$ matrix, Eq.~\eqref{eq:unitarity_of_U} is a consequence of the unitarity of the matrix that diagonalizes neutral lepton masses, 
Eq.~\eqref{eq:more_equations} are a consequence of Eqs.~\eqref{eq:definition_C} and \eqref{eq:unitarity_of_U}, 
and the equalities in Eq.~\eqref{eq:seesaw_relations_reloaded} are the seesaw relation written in terms of $\mathcal{U}$ and $\mathcal{C}$.

It is important to state that all of these relations are true at all orders of $\Theta$, or whether or not we have an $L$ conserving symmetry. 

In the $L$ conserving case with $\mathcal{N} = 2$, we have that $\mathcal{C}$ is equal to
\begin{align}
    \mathcal{C} \simeq \begin{pmatrix}
        \mathbf{V}^\dagger\left(\mathbf{I} - \mathbf{\Theta} \mathbf{\Theta}^\dagger\right)\mathbf{V} & \mathbf{V}^\dagger \mathbf{\Theta} \\
        \mathbf{\Theta}^\dagger \mathbf{V} & \mathbf{\Theta}^\dagger \mathbf{\Theta}
    \end{pmatrix}\,.
\end{align}
The block composed of $\mathbf{\Theta}^\dagger \mathbf{\Theta}$ simplifies to
\begin{align}
    \mathbf{\Theta}^\dagger \mathbf{\Theta} = \frac{U_{\mathrm{tot}}^2}{2} \begin{pmatrix}
        1 & i \\ -i & 1
    \end{pmatrix}\,.
\end{align}

\section{Definition of loop functions} \label{app:loop_formulas}
\subsection{Definitions of $F$ and $G$ functions}
\label{app:definitions_f_and_g}
The $F$ and $G$ functions that appear in the amplitudes of Eqs.~\eqref{eq:monopole_amplitude}--\eqref{eq:down_box_amplitude} depend on the following loop functions
\begin{align}
     F_\gamma(x) &= \frac{7 x^3 - x^2 - 12x}{12(1-x)^3} - \frac{x^4 -
    10x^3 + 12x^2}{6(1-x)^4}\log x\,, \\
  G_\gamma(x) &= -\frac{x(2x^2 + 5x - 1)}{4(1-x)^3} -
  \frac{3x^3}{2(1-x)^4}\log x\,, \\
  F_Z(x) &= -\frac{5 x}{2(1 - x)} - \frac{5x^2}{2(1-x)^2}\log x\,, \\
  G_Z(x,y) &= -\frac{1}{2(x-y)}\left[\frac{x^2(1-y)}{1-x}\log x -
  \frac{y^2(1-x)}{1-y}\log y \right]\,,\\
  H_Z(x,y) &= \frac{\sqrt{xy}}{4(x-y)}\left[\frac{x^2 - 4x}{1 -
    x}\log x - \frac{y^2 - 4y}{1 - y}\log
  y\ \right]\,, \\
  F_\mathrm{box}(x, y) &= 
  \begin{aligned}[t]
    &\frac{1}{x-y}\left\{\left(4 + \frac{xy}{4}\right)\left[\frac{1}{1-x} + \frac{x^2\,\log x}{(1-x)^2} - \frac{1}{1-y} - \frac{y^2\,\log y}{(1-y)^2}\right]\right. \\ 
    &\left. -2xy\left[\frac{1}{1-x} + \frac{x\,\log x}{(1-x)^2}
     - \frac{1}{1-y} - \frac{y\,\log y}{(1-y)^2}
    \right]\right\}\,,
  \end{aligned} \\
  F_{X\mathrm{box}}(x, y) &= 
  \begin{aligned}[t]
    &-\frac{1}{x-y}\left\{\left(1 + \frac{xy}{4}\right)\left[\frac{1}{1-x} + \frac{x^2\,\log x}{(1-x)^2} -\frac{1}{1-y} - \frac{y^2\,\log y}{(1-y)^2}\right]\right. \\
    &\left. -2xy\left[\frac{1}{1-x} + \frac{x\,\log x}{(1-x)^2}
    - \frac{1}{1-y} - \frac{y\,\log y}{(1-y)^2}
    \right]\right\}\,.  
  \end{aligned}
\end{align}
For the convenience of the reader, we present some important limits of these loop functions
\begin{align}
    \label{eq:loop_limits_1}
    G_Z(x, x) &= -\frac{x}{2} - \frac{x \log x}{x - 1}\,, & G_Z(x, 0) &= -\frac{x \log x}{2(x - 1)}\,, \\
    F_\mathrm{box}(x,0) &= \frac{4(1 - x + x\log x)}{(x - 1)^2} & 
    F_{X\mathrm{box}}(x,0) &= -\frac{1 - x + x\log x}{(x - 1)^2} \\ 
    F_\gamma(x) &\xrightarrow{x \to 0} -x\,, & 
    F_\gamma(x) &\xrightarrow{x \to \infty} -\frac{7}{12} - \frac{1}{6} \log x\,, \\
    G_\gamma(x) &\xrightarrow{x \to 0} \frac{1}{4} x\,, &
    G_\gamma(x) &\xrightarrow{x \to \infty} \frac{1}{2}\,, \\
    F_Z(x) &\xrightarrow{x \to 0} -\frac{5}{2} x\,, &
    F_Z(x) &\xrightarrow{x \to \infty} \frac{5}{2} - \frac{5}{2}\log x\,, \\
    G_Z(x,x) &\xrightarrow{x \to 0} -\frac{1}{2}\left(1 + 2 \log x \right)x\,, &
    G_Z(x,x) &\xrightarrow{x \to \infty} -\frac{x}{2} + \log x\,, \\
    G_Z(x,0) &\xrightarrow{x \to 0} -\frac{1}{2} x \log x \,,&
    G_Z(x,0) &\xrightarrow{x \to \infty} \frac{1}{2}\,\log x\,, \\
    F_\mathrm{box}(x,0) &\xrightarrow{x \to 0} 4\,, &
    F_\mathrm{box}(x,0) &\xrightarrow{x \to \infty} 0\,, \\
    \label{eq:loop_limits_9}
    F_{X\mathrm{box}}(x,0) &\xrightarrow{x \to 0} 1\,, &
    F_\mathrm{box}(x,0) &\xrightarrow{x \to \infty} 0\,.
\end{align}

All these functions should be plugged into
\begin{align}
    \label{eq:photon_loop}
    F_{\gamma}^{\beta\alpha} &= \sum_{k = 1}^{3 + \mathcal{N}} \mathcal{U}_{\alpha k}^{\phantom{\ast}}\,\mathcal{U}_{\beta k}^{\ast}\,F_{\gamma} (x_k)\,, \\
    \label{eq:dipole_loop}
    G_{\gamma}^{\beta\alpha} &= \sum_{k = 1}^{3 + \mathcal{N}} \mathcal{U}_{\alpha k}^{\phantom{\ast}}\,\mathcal{U}_{\beta k}^{\ast}\,G_{\gamma} (x_k)\,, \\
    F_Z^{\beta\alpha} &= \sum_{k,l = 1}^{3 + \mathcal{N}} \mathcal{U}_{\alpha k}^{\phantom{\ast}}\,\mathcal{U}_{\beta l}^{\ast}
    \label{eq:Z_loop}
    \left[\delta_{kl} F_Z(x_k) + \mathcal{C}_{kl}^{\phantom{\ast}} \,G_Z(x_k, x_l) + \mathcal{C}_{kl}^{\ast}\,H_Z(x_k,x_l)\right]\,,\\
    \label{eq:u_loop}
    F_{\mathrm{u},box}^{ji\beta\alpha} &= \sum_{d_l = d, s, b}\sum_{k = 1}^{3 + \mathcal{N}} \mathcal{U}_{\alpha k}^{\phantom{\ast}}\,\mathcal{U}_{\beta k}^{\ast}\,\tilde{V}_{u_j d_l}\,\tilde{V}_{u_i d_l}^{\ast}\,F_{\mathrm{box}}(x_k, x_{d_l})\,, \\
    \label{eq:down_loop}
    F_{\mathrm{d},box}^{ji\beta\alpha} &= \sum_{u_l = u, c, t}\sum_{k = 1}^{3 + \mathcal{N}} \mathcal{U}_{\alpha k}^{\phantom{\ast}}\,\mathcal{U}_{\beta k}^{\ast}\,\tilde{V}_{u_l d_i}\,\tilde{V}_{u_l d_j}^{\ast}\,F_{X\mathrm{box}}(x_k, x_{u_l})\,,
\end{align}
where $\tilde{V}_{u_i d_j}$ are CKM matrix elements, $x_k = m_k^2/M_W^2$ is the mass of neutral leptons divided by the $W$ boson mass, and $x_{u_i} = m_{u_i}^2/M_W^2, x_{d_i} = m_{d_i}^2/M_W^2$.

For the $L$ conserving case with only 2 HNLs, we can simplify Eqs.~\eqref{eq:photon_loop} -- \eqref{eq:Z_loop} to
\begin{align}
    F_{\gamma}^{\beta\alpha} &= \Lambda_{\alpha\beta}\, U_{\mathrm{tot}}^2\,F_\gamma(x_N)\,, \\
    G_{\gamma}^{\beta\alpha} &= \Lambda_{\alpha\beta}\, U_{\mathrm{tot}}^2\,G_\gamma(x_N)\,, \\
    F_Z^{\beta\alpha} &= 
    \Lambda_{\alpha\beta}\,U_{\mathrm{tot}}^2 \left\lbrace F_Z(x_N) + 2\,G_Z(x_N,0) + U_{\mathrm{tot}}^2 \left[G_Z(x_N, x_N) - 2\,G_Z(x_N,0) \right]\right\rbrace\,,
\end{align}
where we considered active neutrinos to be massless, and where we took advantage of all the formulas introduced in Appendix~\ref{app:unitarity_seesaw_formulas}. 
Notice that the part proportional to $H_Z$ in Eq.~\eqref{eq:Z_loop} disappeared, which is exactly zero for the $L$ conserving case.

The loop boxes in Eqs.~\eqref{eq:u_loop} and \eqref{eq:down_loop} depend on quark masses. If we neglect the masses of all quarks, except for the $t$ quark, then the loop function simplifies to
\begin{align}
    \label{eq:up_box_L_symmetry}
    F_{\mathrm{u},box}^{ji\beta\alpha} &= \Lambda_{\alpha\beta}\,U_{\mathrm{tot}}^2\,\delta_{u_i u_j}\left[F_\mathrm{box}(x_N, 0) - F_\mathrm{box}(0, 0) \right]\,,\\
    \label{eq:down_box_L_symmetry}
    F_{\mathrm{d},box}^{ji\beta\alpha} &= 
    \begin{aligned}[t]
        &\Lambda_{\alpha\beta}\,U_{\mathrm{tot}}^2 \left\lbrace \delta_{d_i d_j}\left[F_{X\mathrm{box}}(x_N, 0) - F_{X\mathrm{box}}(0, 0) \right] \right. \\
        &+ \left.\tilde{V}_{t d_i}\,\tilde{V}_{t d_j}^\ast \left[F_{X\mathrm{box}}(x_N, x_t) - F_{X\mathrm{box}}(x_N,0) - F_{X\mathrm{box}}(0, x_t) + F_{X\mathrm{box}}(0, 0) \right]\right\rbrace\,.
    \end{aligned}
\end{align}
We can see in both Eq.~\eqref{eq:up_box_L_symmetry} and \eqref{eq:down_box_L_symmetry} the usual GIM suppression \cite{Glashow:1970gm} one would expect from flavor-changing neutral currents in the quark sector from weak interactions. 
Indeed, the unitarity of the CKM matrix and the lightness of quarks compared to the electroweak scale kill quark flavor-changing processes.
An exception can be seen in Eq.~\eqref{eq:down_box_L_symmetry}, where the mass of the top quark does not completely suppress flavor-changing currents, particularly in interactions involving a bottom quark. 

\subsection{Definitions of $\mathcal{F}$ and $\mathcal{G}$ functions}
\label{app:defintions_of_F_and_G}
In this subsection we will define the $\mathcal{F}$ and $\mathcal{G}$ functions that in the amplitudes in Eqs.~\eqref{eq:monopole_amplitude} -- \eqref{eq:down_box_amplitude}, and in the branching ratios in Eqs.~\eqref{eq:decay_to_pseudoscalar}--\eqref{eq:decay_two_mesons}. 
All of these formulas can easily be derived from the quark currents in Table~\ref{tab:bound_states_mesons}
\begin{align}
    \mathcal{F}_\pi^{\beta\alpha} &= \frac{f_{\pi^0}}{4\sqrt{2}} \left( 2 F_Z^{\beta \alpha} + F_{\mathrm{u,box}}^{uu\beta\alpha} + F_{\mathrm{d,box}}^{dd\beta\alpha} \right)\,, \\
    \mathcal{F}_\eta^{\beta\alpha} &= \frac{f_8\,c_8}{4\sqrt{6}} 
        \left(-2 F_Z^{\beta \alpha} - F_{\mathrm{u,box}}^{uu\beta\alpha} + F_{\mathrm{d,box}}^{dd\beta\alpha} - 2 F_{\mathrm{d,box}}^{ss\beta\alpha} \right) \\ \nonumber
        &\quad- \frac{f_0\,s_0}{4\sqrt{3}} 
        \left(F_Z^{\beta \alpha} - F_{\mathrm{u,box}}^{uu\beta\alpha} + F_{\mathrm{d,box}}^{dd\beta\alpha} + F_{\mathrm{d,box}}^{ss\beta\alpha} \right)\,,\\
    \mathcal{F}_{\eta^\prime}^{\beta\alpha} &= 
    \frac{f_8\,s_8}{4\sqrt{6}} 
    \left(-2 F_Z^{\beta \alpha} - F_{\mathrm{u,box}}^{uu\beta\alpha} + F_{\mathrm{d,box}}^{dd\beta\alpha} - 2 F_{\mathrm{d,box}}^{ss\beta\alpha} \right) + \\
    &\quad + \frac{f_0\,c_0}{4\sqrt{3}} \left(F_Z^{\beta \alpha} - F_{\mathrm{u,box}}^{uu\beta\alpha} + F_{\mathrm{d,box}}^{dd\beta\alpha} + F_{\mathrm{d,box}}^{ss\beta\alpha} \right)\,, \\
    \mathcal{F}_{\rho}^{\beta\alpha} &= \frac{f_\rho\,m_\rho}{\sqrt{2}} 
    \left[ s_W^2 F_\gamma^{\beta\alpha} + \left(\frac{1}{2} - s_W^2 \right) F_Z^{\beta\alpha} + \frac{1}{4} \left(F_{\mathrm{u,box}}^{uu\beta\alpha} + F_{\mathrm{d,box}}^{dd\beta\alpha} \right) \right]\,, \\
    \mathcal{F}_{\omega}^{\beta\alpha} &= \frac{f_\omega\,m_\omega}{\sqrt{2}} 
    \left[\frac{1}{3}s_W^2 \left( F_\gamma^{\beta\alpha} - F_Z^{\beta\alpha} \right) + \frac{1}{4} \left(F_{\mathrm{u,box}}^{uu\beta\alpha} - F_{\mathrm{d,box}}^{dd\beta\alpha} \right) \right]\,, \\
    \mathcal{F}_{\phi}^{\beta\alpha} &= f_\phi\,m_\phi 
    \left[-\frac{1}{3}s_W^2 F_\gamma^{\beta\alpha} + \left(-\frac{1}{4} + \frac{1}{3}s_W^2 \right) F_Z^{\beta\alpha} - \frac{1}{4} F_{\mathrm{d,box}}^{ss\beta\alpha} \right]\,, \\
    \mathcal{F}_{J/\Psi}^{\beta\alpha} &= f_{J/\psi}\,m_{J/\psi}
    \left[\frac{2}{3}s_W^2 F_\gamma^{\beta\alpha} + \left(\frac{1}{4} - \frac{2}{3}s_W^2 \right) F_Z^{\beta\alpha} + \frac{1}{4} F_{\mathrm{u,box}}^{cc\beta\alpha} \right]\,, \\
    \mathcal{F}_{\Upsilon}^{\beta\alpha} &= f_{\Upsilon}\,m_{\Upsilon}
    \left[-\frac{1}{3}s_W^2 F_\gamma^{\beta\alpha} + \left(-\frac{1}{4} + \frac{1}{3}s_W^2 \right) F_Z^{\beta\alpha} - \frac{1}{4} F_{\mathrm{d,box}}^{bb\beta\alpha} \right]\,, \\
    \mathcal{F}_{\pi \pi}^{\beta\alpha} &= F_{\pi\pi}(s) \left[s_W^2 F_\gamma^{\beta\alpha} + \left(\frac{1}{2} - s_W^2 \right) F_Z^{\beta\alpha} + \frac{1}{4} \left(F_{\mathrm{u,box}}^{uu\beta\alpha} + F_{\mathrm{d,box}}^{dd\beta\alpha} \right) \right]\,, \\
    \mathcal{F}_{K^+K^-}^{\beta\alpha} &= 
        F_{KK}^{(3)}(s)\left[s_W^2 F_\gamma^{\beta\alpha} + \left(\frac{1}{2} - s_W^2 \right) F_Z^{\beta\alpha} + \frac{1}{4} \left(F_{\mathrm{u,box}}^{uu\beta\alpha} + F_{\mathrm{d,box}}^{dd\beta\alpha} \right) \right] \\ \nonumber
        &\quad + F_{KK}^{(8)}(s)\left[s_W^2 F_\gamma^{\beta\alpha} + \left(\frac{1}{2} - s_W^2 \right) F_Z^{\beta\alpha} + \frac{1}{4} \left(F_{\mathrm{u,box}}^{uu\beta\alpha} - F_{\mathrm{d,box}}^{dd\beta\alpha} + 2F_{\mathrm{d,box}}^{ss\beta\alpha} \right) \right] \\ \nonumber
        &\quad + F_{KK}^{(0)}(s)\left[-\frac{1}{4} F_Z^{\beta\alpha} + \frac{1}{4} \left(F_{\mathrm{u,box}}^{uu\beta\alpha} - F_{\mathrm{d,box}}^{dd\beta\alpha} - F_{\mathrm{d,box}}^{ss\beta\alpha} \right) \right] \,,\\[2ex]
    \mathcal{F}_{K_S\,K_S}^{\beta\alpha} &=
        -\frac{F_{KK}^{(3)}(s)}{\sqrt{2}}\left[s_W^2 F_\gamma^{\beta\alpha} + \left(\frac{1}{2} - s_W^2 \right) F_Z^{\beta\alpha} + \frac{1}{4} \left(F_{\mathrm{u,box}}^{uu\beta\alpha} + F_{\mathrm{d,box}}^{dd\beta\alpha} \right) \right] \\ \nonumber
        & \quad + \frac{F_{KK}^{(8)}(s)}{\sqrt{2}}\left[s_W^2 F_\gamma^{\beta\alpha} + \left(\frac{1}{2} - s_W^2 \right) F_Z^{\beta\alpha} + \frac{1}{4} \left(F_{\mathrm{u,box}}^{uu\beta\alpha} - F_{\mathrm{d,box}}^{dd\beta\alpha} + 2F_{\mathrm{d,box}}^{ss\beta\alpha} \right) \right] \\ \nonumber
        & \quad + \frac{F_{KK}^{(0)}(s)}{\sqrt{2}}\left[-\frac{1}{4} F_Z^{\beta\alpha} + \frac{1}{4} \left(F_{\mathrm{u,box}}^{uu\beta\alpha} - F_{\mathrm{d,box}}^{dd\beta\alpha} - F_{\mathrm{d,box}}^{ss\beta\alpha} \right) \right] \,,\\[2ex]
    \mathcal{G}_{\rho}^{\beta\alpha} &= \frac{f_{\rho}\,m_\rho}{\sqrt{2}} s_W^2 G_{\gamma}^{\beta\alpha}\,, \\
    \mathcal{G}_{\omega}^{\beta\alpha} &= \frac{f_{\omega}\,m_\omega}{3\sqrt{2}} s_W^2 G_{\gamma}^{\beta\alpha}\,,\\
    \mathcal{G}_{\phi}^{\beta\alpha} &= -\frac{f_{\phi}\,m_\phi}{3} s_W^2 G_{\gamma}^{\beta\alpha}\,,\\
    \mathcal{G}_{J/\Psi}^{\beta\alpha} &= \frac{2 f_{J/\Psi}\,m_{J/\Psi}}{3} s_W^2 G_{\gamma}^{\beta\alpha}\,,\\
    \mathcal{G}_{\Upsilon}^{\beta\alpha} &= -\frac{f_{\Upsilon}\,m_{\Upsilon}}{3} s_W^2 G_{\gamma}^{\beta\alpha}\,,\\
    \mathcal{G}_{\pi\pi}^{\beta\alpha} &= F_{\pi\pi}(s)\,s_W^2 G_{\gamma}^{\beta\alpha}\,,\\
    \mathcal{G}_{K^+K^-}^{\beta\alpha} &= \left[F_{KK}^{(3)}(s) + F_{KK}^{(8)}(s) \right] s_W^2 G_{\gamma}^{\beta\alpha}\,, \\
    \mathcal{G}_{K_S\,K_S}^{\beta\alpha} &= \frac{1}{\sqrt{2}}\left[-F_{KK}^{(3)}(s) + F_{KK}^{(8)}(s) \right] s_W^2 G_{\gamma}^{\beta\alpha}\,.
\end{align}

The values of all the decay constants are shown in Table~\ref{tab:bound_states_mesons}, and we shall show the parametrization of the form factors in Appendix~\ref{app:form_factors}.

\section{General branching ratio formulas}
\label{app:branching_ratios_general}
\begin{align}
    \label{eq:decay_to_pseudoscalar}
    \mathrm{Br}(\ell_\beta \to \ell_\alpha\,P) &= \frac{\alpha_W^4\,\lambda^{\frac{1}{2}}(m_\beta^2, m_P^2, m_\alpha^2)}{128\pi\,M_W^4\,m_\beta^3\,\Gamma_\beta}\,\abs{\mathcal{F}_P^{\beta\alpha}}^2 \left[\left(m_\beta^2 - m_\alpha^2 \right)^2 - m_P^2\,\left(m_\beta^2 + m_\alpha^2 \right)\right]\,, \\
    \label{eq:pseudoscalar_decay}
    \mathrm{Br}(P \to \ell_\alpha \ell_\beta) &= \frac{\alpha_W^4\,\lambda^{\frac{1}{2}}(m_P^2, m_\alpha^2, m_\beta^2)}{64\pi\,M_W^4\,m_P^3\,\Gamma_P}\,\abs{\mathcal{F}_P^{\beta\alpha}}^2 \left[m_P^2\,\left(m_\alpha^2 + m_\beta^2 \right) -  \left(m_\alpha^2 - m_\beta^2 \right)^2\right]\,,
\end{align}

\begin{align}
    \label{eq:vector_decay}
    \mathrm{Br}(V \to \ell_\alpha \ell_\beta) &= 
    \begin{aligned}[t]
        &\frac{\alpha_W^4\,\lambda^{\frac{1}{2}}(m_V^2, m_\alpha^2, m_\beta^2)}{192\pi\,M_W^4\,m_V^5\,\Gamma_V} \left[\abs{\mathcal{F}_V^{\beta\alpha}}^2\,(2 m_V^4 - m_V^2\left(m_\alpha^2 + m_\beta^2 \right) - (m_\alpha^2 - m_\beta^2)^2) \right. \\
        & + 3\,\Re[\mathcal{F}^{\beta\alpha}_V\,\mathcal{G}_V^{\beta\alpha\ast}] \left[-m_V^2 \left( m_\alpha^2 + m_\beta^2 \right)^2 + \left( m_\alpha^2 - m_\beta^2 \right)^2\right]\\
        &+ \left.\abs{\mathcal{G}_V^{\beta\alpha}}^2 \frac{m_V^4 \left(m_\alpha^2 + m_\beta^2 \right) + m_V^2 \left(m_\alpha^4 + 14 m_\alpha^2 m_\beta^2   + m_\beta^4 \right) - 2 \left( m_\alpha^2 + m_\beta^2 \right) \left( m_\alpha^2 - m_\beta^2 \right)^2}{m_V^2}  \right]\,,
    \end{aligned}
    \\
    \label{eq:decay_to_vector}
    \mathrm{Br}(\ell_\beta \to \ell_\alpha V) &= 
    \begin{aligned}[t]
    &\frac{\alpha_W^4 \sqrt{\lambda(m_\beta^2, m_V^2, m_\alpha^2)}}{128\pi\,M_W^4\,m_V^2\,m_\beta^3\, \Gamma_\beta} \left[\abs{\mathcal{F}_V^{\beta\alpha}}^2\, ((m_\beta^2 - m_\alpha^2)^2 + m_V^2\,(m_\beta^2 + m_\alpha^2) - 2 m_V^4)\right. \\
    &+ 6 \Re[\mathcal{F}_V^{\alpha\beta}\,\mathcal{G}_V^{\alpha\beta\ast}] ((m_\beta^2 - m_\alpha^2)^2 - m_V^2(m_\beta^2 + m_\alpha^2)) \\
    &\left.+ \abs{\mathcal{G}_V^{\beta\alpha}}^2 \frac{2 \left( m_\alpha^2 + m_\beta^2 \right) \left( m_\alpha^2 - m_\beta^2 \right)^2 - m_V^4 \left(m_\alpha^2 + m_\beta^2 \right) - m_V^2 \left(m_\alpha^4 + 14 m_\alpha^2 m_\beta^2   + m_\beta^4 \right)}{ m_V^2}\right]\,,
    \end{aligned}
    \\
    \label{eq:decay_two_mesons}
    \mathrm{Br}(\ell_\beta \to \ell_\alpha P P) &= 
    \begin{aligned}[t]
    &\frac{\alpha_W^2}{6144\pi\,M_W^4\,m_\beta^3\,\Gamma_\beta}\int_{4 m_P^2}^{(m_\beta - m_\alpha)^2} \dd s \sqrt{\lambda(s, m_\beta^2, m_\alpha^2) \lambda(s, m_P^2, m_P^2)}\,\frac{s-4m_P^2}{s^2} \\
    &\left[\abs{\mathcal{F}_{PP}^{\beta\alpha}}^2\,((m_\beta^2 - m_\alpha^2)^2 + s\,(m_\alpha^2 + m_\beta^2) -  2 s^2) \right. \\
    &+ 6 \Re[\mathcal{F}_{PP}^{\beta\alpha}\,\mathcal{G}_{PP}^{\beta\alpha\ast}] ((m_\beta^2 - m_\alpha^2)^2 - s (m_\beta^2 + m_\alpha^2)) \\
    &\left.+ \abs{\mathcal{G}_{PP}^{\beta\alpha}}^2\,\frac{2(m_\beta^2 - m_\alpha^2)^2 (m_\beta^2 + m_\alpha^2) - s^2(m_\beta^2 + m_\alpha^2) - s(m_\beta^4 + 14 m_\alpha^2\,m_\beta^2 + m_\alpha^4)}{s}\right]
    \end{aligned}\,.
\end{align}

\section{Parametrization of form factors}
\label{app:form_factors}
As we highlighted in the main text, we define of the pion and kaon systems as
\begin{align}
    \mel{\pi^+(p_{\pi^+})\,\pi^-(p_{\pi^-})}{\frac{1}{2}(\bar{u}\gamma^\mu u - \bar{d}\gamma^\mu d )}{0} &= F_{\pi\pi}(s)\,(p_{\pi^+} - p_{\pi^-})^\mu\,, \\
    \label{eq:isovector_kaons}
    \mel{K^+(p_{K^+})\,K^-(p_{K^-})}{\frac{1}{2}(\bar{u}\gamma^\mu u - \bar{d}\gamma^\mu d )}{0} &= F_{KK}^{(3)}(s)\,(p_{K^+} - p_{K^-})^\mu\,, \\
    \mel{K^+(p_{K^+})\,K^-(p_{K^-})}{\frac{1}{6}(\bar{u}\gamma^\mu u + \bar{d}\gamma^\mu d - 2 \bar{s} \gamma^\mu s )}{0} &= F_{KK}^{(8)}(s)\,(p_{K^+} - p_{K^-})^\mu\,, \\
    \mel{K^+(p_{K^+})\,K^-(p_{K^-})}{\frac{1}{3}(\bar{u}\gamma^\mu u + \bar{d}\gamma^\mu d + \bar{s} \gamma^\mu s)}{0} &= F_{KK}^{(0)}(s)\,(p_{K^+} - p_{K^-})^\mu\,,
\end{align}
with final state $\ket{K^0 \bar{K}^0}$ having also the same definition as with charged kaon, with an additional minus sign in the isovector component in Eq.~\eqref{eq:isovector_kaons} ~\cite{Bruch:2004py, Beloborodov:2019uql}.

We used the vector meson dominated (VMD) models for the parametrization of different form factors \cite{Sakurai:1960ju}, where we assume that interactions that involve hadrons are mediated by a sum of different vector bound states, such as $\rho, \omega$, and $\phi$-like resonances. 
Accordingly, we parametrize form factors as a sum of different Breit-Wigner (BW) functions, for a form factor, this looks as
\begin{align}
    F(s) = \sum_i c_i\, \mathrm{BW}(s, m_{V_i}, \Gamma_{V_i})\,,
\end{align}
where $c_i$ are normalization constants, that in general can be complex, that ensure that $F(s = 0) = 1$. 
Dual-QCD in the limit of $N_C = \infty$ predicts that there should be an infinite number of these resonances \cite{Witten:1979kh, Bruch:2004py}, but phenomenologically the most relevant ones are the first ones.

We parametrize the $\rho, \omega$, and $\phi$ resonances with different BW functions, just as it was used in \cite{Czyz:2010hj}. 
For $\rho$, we used the Gounaris-Sakurai (GS) BW function \cite{Gounaris:1968mw}, for the $\omega$ resonances a constant BW, and for $\phi$ resonances we used a Kühn-Santamaría (KS) BW function \cite{Kuhn:1990ad}. Following the parametrization of \cite{Kuhn:1990ad, Czyz:2010hj}, the BW functions are defined as 
\begin{align}
    \mathrm{BW}^{\mathrm{GS}}(s, m_V, \Gamma_V) &= \frac{m_V^2 + h(0, m_V, \Gamma_V)}{m_V^2 - s + h(s, m_V, \Gamma_V) - i \sqrt{s}\,\tilde{\Gamma}_V(s, m_V, \Gamma_V)}\,, \\
    \mathrm{BW}^{\mathrm{KS}}(s, m_V, \Gamma_V) &= \frac{m_V^2}{m_V^2 - s - i \sqrt{s}\,\tilde{\Gamma}_V(s, m_V, \Gamma_V)}\,, \\
    \mathrm{BW}^{\mathrm{const}}(s, m_V, \Gamma_V) &= \frac{m_V^2}{m_V^2 - s - i \sqrt{s}\,\Gamma_V}\,,
\end{align}
where the auxiliary functions are defined as
\begin{align}
    &\tilde{\Gamma}(s, m_V, \Gamma_V) = \frac{m_V^2}{s}\left(\frac{s - 4\,m_P^2}{m_V^2 - 4\,m_P^2}\right)^{3/2}\Gamma_V\,, \\
    &h(s, m_V, \Gamma_V) = \hat{h}_V(s) - \hat{h}_V(m_V^2) - (s - m_V^2) \left.\dv{s}\hat{h}_V(s)\right|_{s \to m_V^2}\,, \\
    &\hat{h}_V(s) = \frac{4\,m_V^2\,\Gamma_V}{\pi (s - 4\,m_P^2)} \left(\frac{s}{4} - m_P^2 \right) \left(1 - \frac{4\,m_P^2}{s} \right)^{1/2} \log(\frac{1 + \left(1 - \frac{4\,m_P^2}{s}\right)^{1/2}}{1 - \left(1 - \frac{4\,m_P^2}{s}\right)^{1/2}})\,,
\end{align}
where $m_P$ should be understood as either the mass of the pion or kaon, depending on the form factor in question. 
The rest of the parameters, $c_V, m_V, \Gamma_V$, are fitted from data from different experiments from different processes.

For $F_{\pi\pi}(s)$, we used the fit from the BaBar collaboration \cite{BaBar:2012bdw}, which was written as a sum of four different $\rho$-like vector mesons
\begin{align}
    F_{\pi\pi}(s) &= \frac{1}{1 + c_{\rho^{\prime}} + c_{\rho^{\prime\prime}} + c_{\rho^{\prime\prime\prime}}}\sum_{V = \rho, \rho^\prime\dots}^{\rho^{\prime\prime\prime}} c_{V} \mathrm{BW}^{\mathrm{GS}}(s, m_{V}, \Gamma_{V})\,.
\end{align}
To account for $\rho-\omega$ mixing, we shift the first resonance
\begin{align}
    \mathrm{BW}^{\mathrm{GS}}(s, m_{\rho}, \Gamma_{\rho}) \to \mathrm{BW}^{\mathrm{GS}}(s, m_{\rho}, \Gamma_{\rho})\,\frac{1 + c_\omega\, \mathrm{BW}^{\mathrm{const}}(s, m_\omega, \Gamma_\omega)}{1 + c_\omega}\,.
\end{align}

The different kaon form factors are written as \cite{Bruch:2004py, Cirigliano:2021img}
\begin{align}
    F_{KK}^{(3)}(s) &= \frac{1}{2} \sum_{V = \rho,  \rho^\prime\dots} c_V\,\mathrm{BW}^{\mathrm{GS}}(s, m_V, \Gamma_V)\,, \\
    F_{KK}^{(8)}(s) &= \frac{1}{6} \sum_{V = \omega,  \omega^\prime\dots} c_V\,\mathrm{BW}^{\mathrm{const}}(s, m_V, \Gamma_V) + \frac{1}{3} \sum_{V = \phi,  \phi^\prime\dots} c_V\,\mathrm{BW}^{\mathrm{KS}}(s, m_V, \Gamma_V)\,, \\
    \label{eq:isosinglet_form_factor}
    F_{KK}^{(0)}(s) &= \frac{1}{3}\sum_{V = \omega,  \omega^\prime\dots} c_V\,\mathrm{BW}^{\mathrm{const}}(s, m_V, \Gamma_V) - \frac{1}{3} \sum_{V = \phi,  \phi^\prime\dots} c_V\,\mathrm{BW}^{\mathrm{KS}}(s, m_V, \Gamma_V)\,.
\end{align}

We used the output from the fit of \cite{Plehn:2019jeo} (see \cite{Czyz:2010hj} for details on the parametrization). 
We did not use the infinite tower of resonances.
Instead, for the $\rho$-like resonances, we used the first six resonances, and for $\omega$ and $\phi$-like, we used the first five. 

The $\phi$-like resonances in the isoscalar ($F_{KK}^{(8)}$) and isosinglet ($F_{KK}^{(0)}$) components are different for neutral and charged kaon final states, due to the isospin-breaking difference between charged and neutral kaon couplings~\cite{Bruch:2004py}. 

The isosinglet component, $F_{KK}^{(0)}(s)$, cannot be extracted directly from data. 
We will follow the steps in \cite{Cirigliano:2021img}, and use the shape in Eq.~\eqref{eq:isosinglet_form_factor}, which is well motivated by theory.
We also verified that the fit done in \cite{Plehn:2019jeo} is consistent with $F_{KK}^{(0)}(0)  = 1$, as expected from chiral perturbation theory.

\newpage
\section{Additional plots}
\label{app:additional_plots} 
\begin{figure}[!h]
    \centering
    \includegraphics[width=\linewidth]{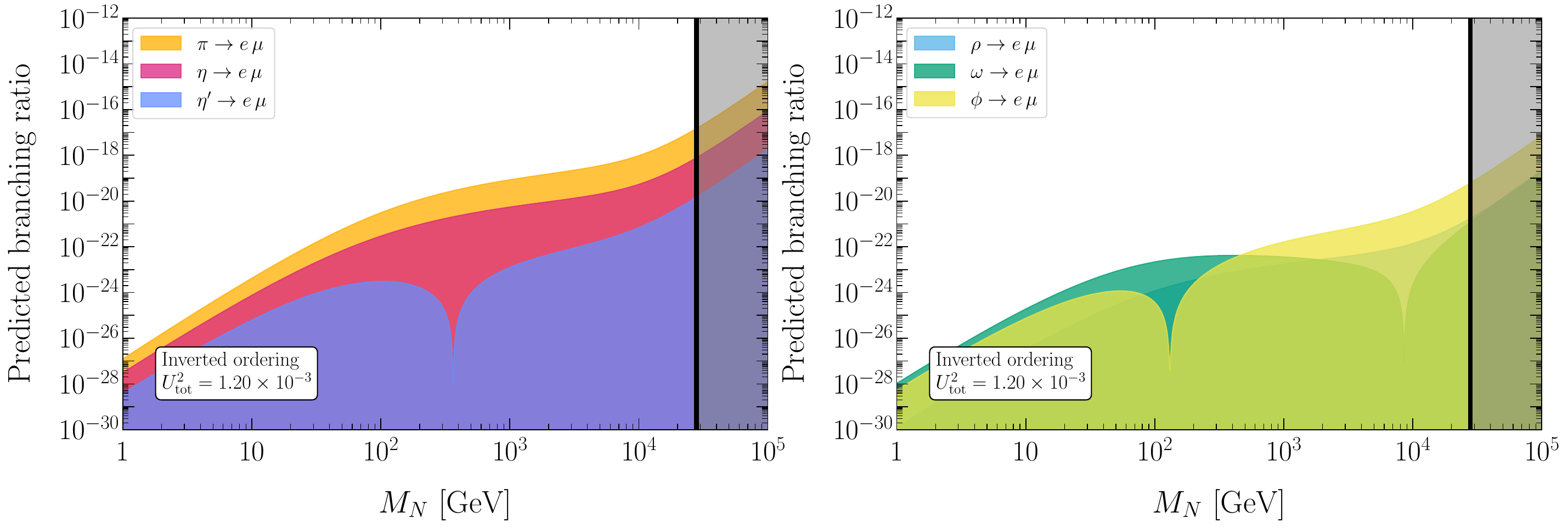}
    \includegraphics[width=\linewidth]{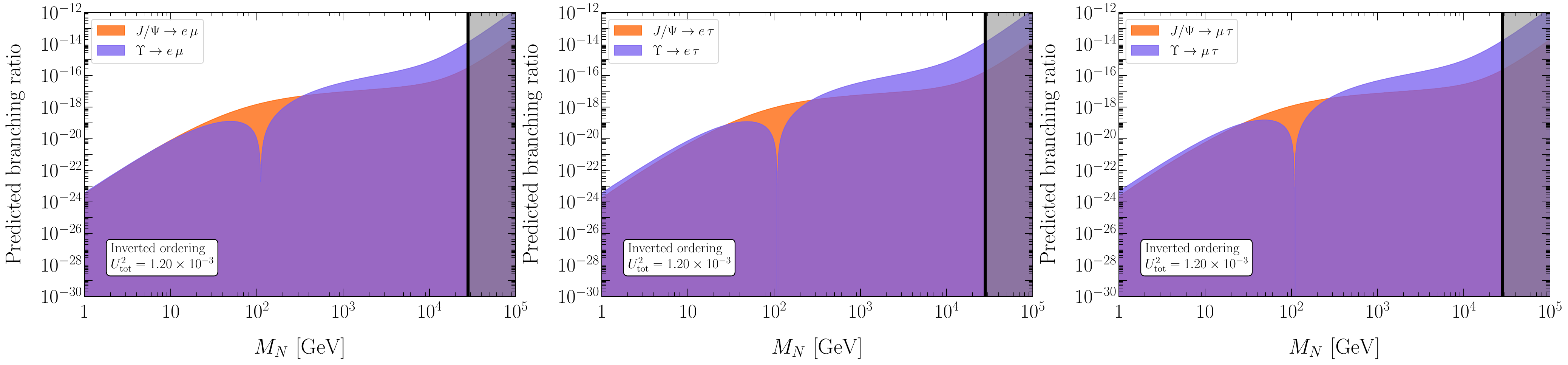}
    \caption{Predictions for light unflavored meson and quarkonia cLFV decays.}
    \label{fig:predictions_meson_decays_IO}
\end{figure}

\begin{figure}[!h]
    \centering
    \includegraphics[width=\linewidth]{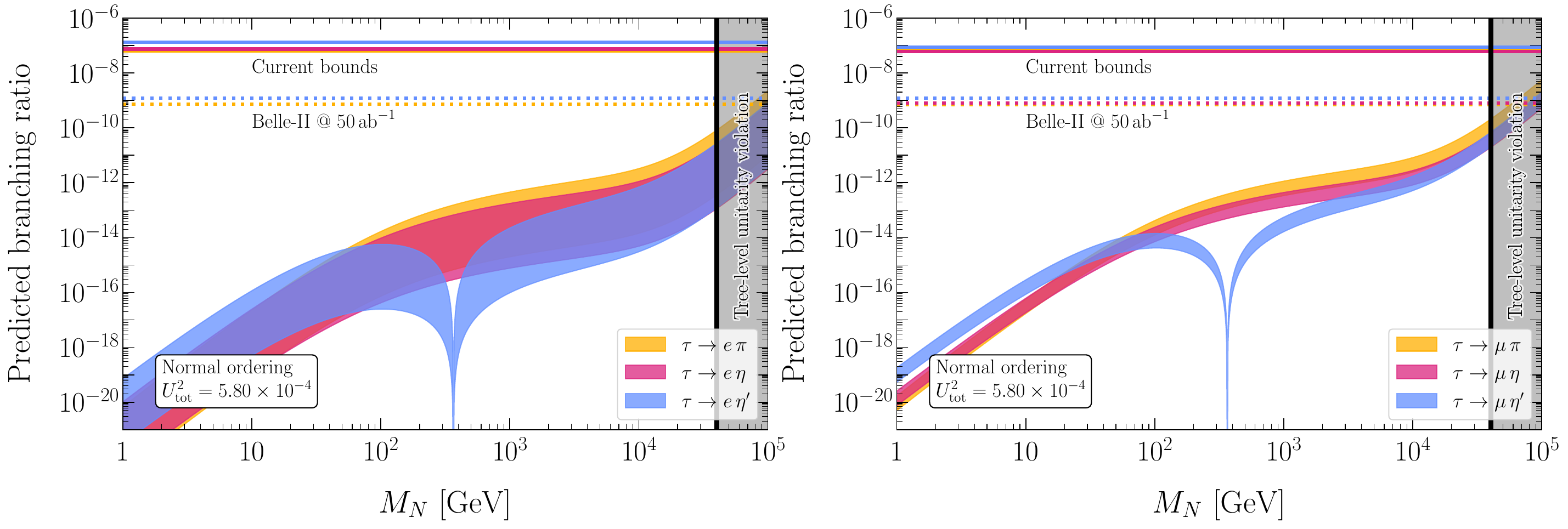}
    \includegraphics[width=\linewidth]{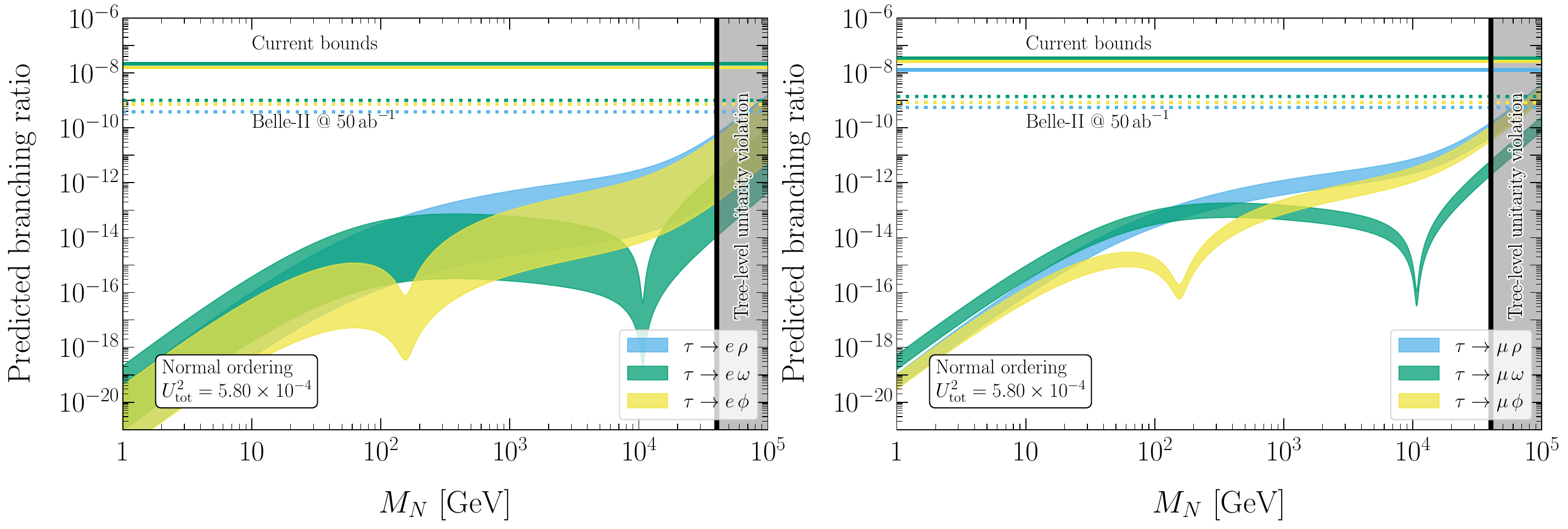}
    \includegraphics[width=\linewidth]{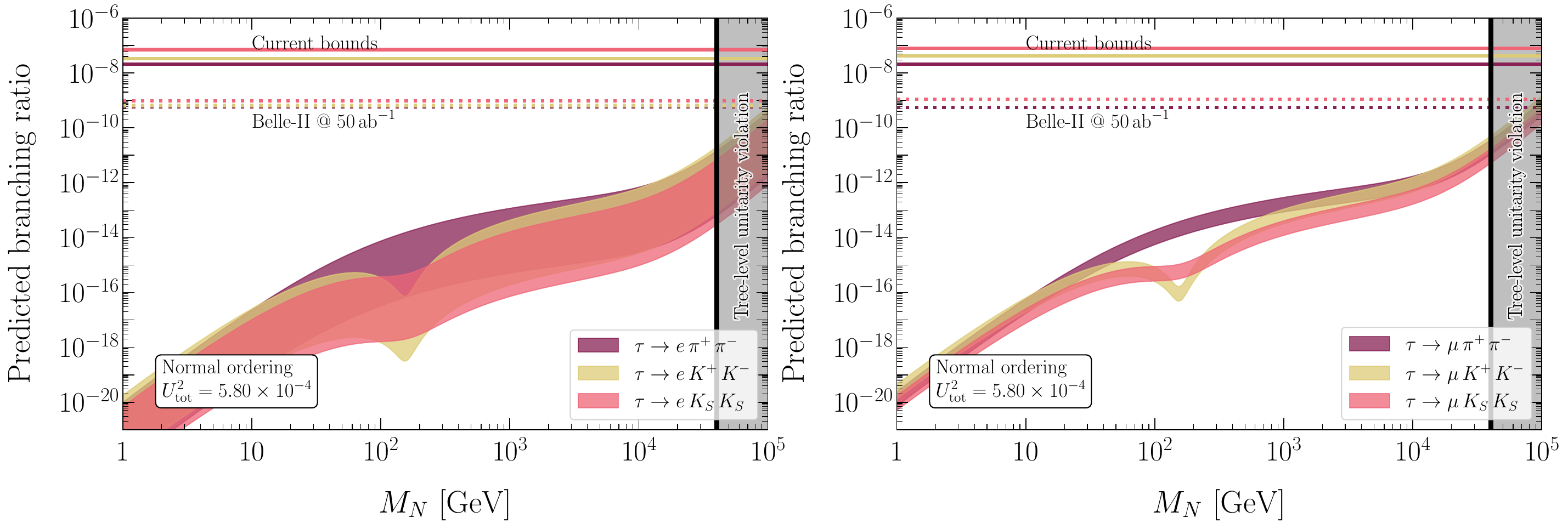}
    \includegraphics[width = \linewidth]{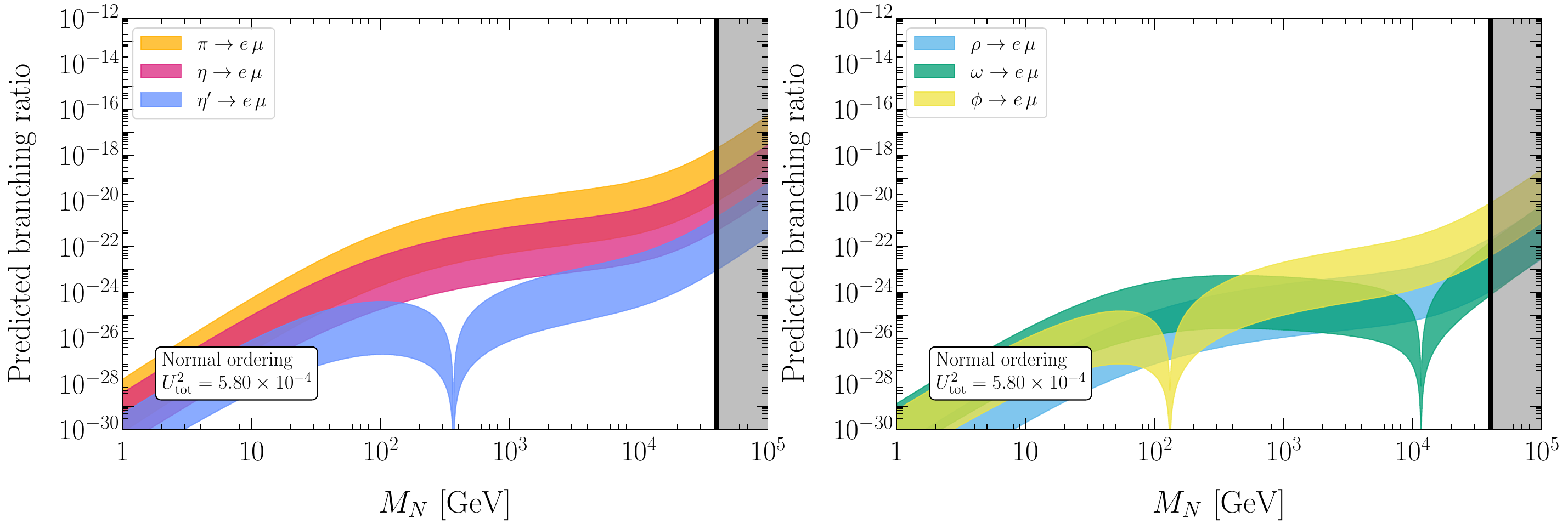}
    \includegraphics[width = \linewidth]{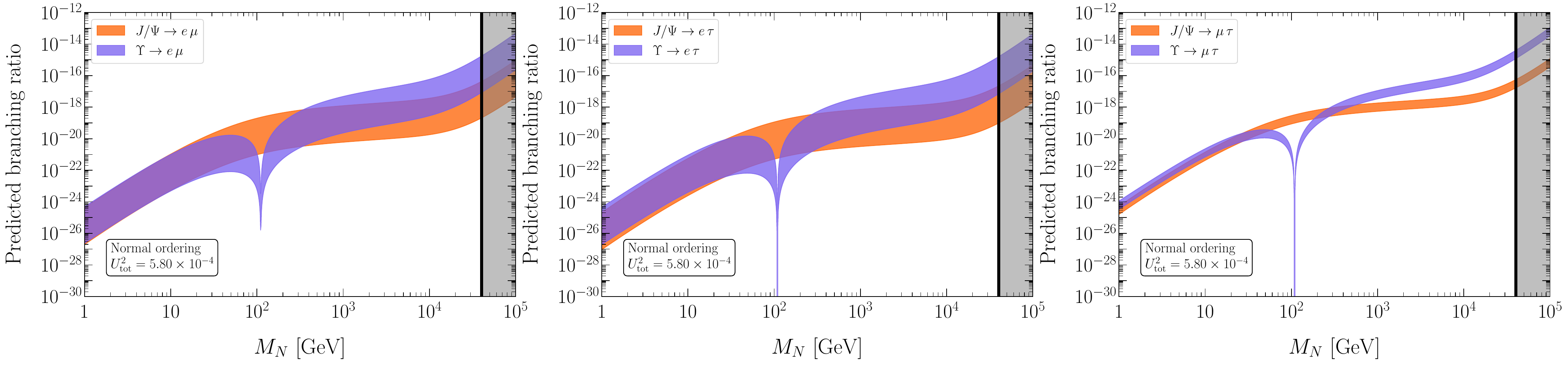}
    \caption{Predictions for cLFV decays with mesons }
    \label{fig:predictions_decays_NO}
\end{figure}

\begin{figure}[!h]
    \centering
    \includegraphics[width = \linewidth]{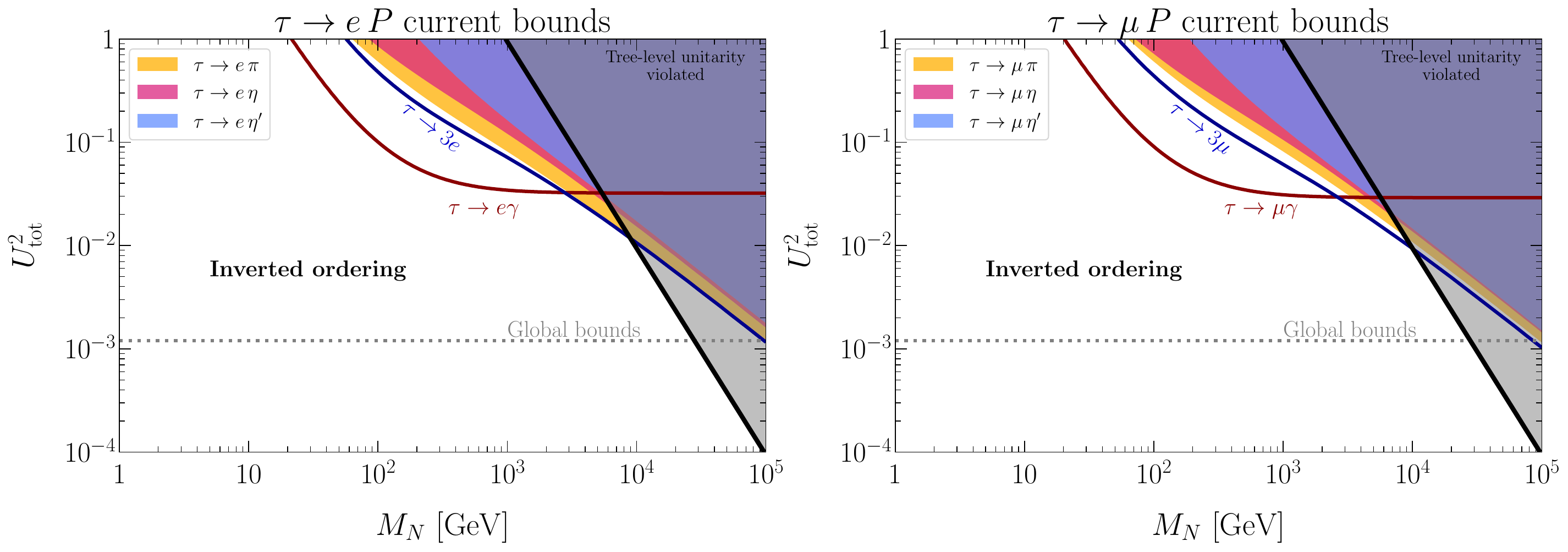}
    \includegraphics[width = \linewidth]{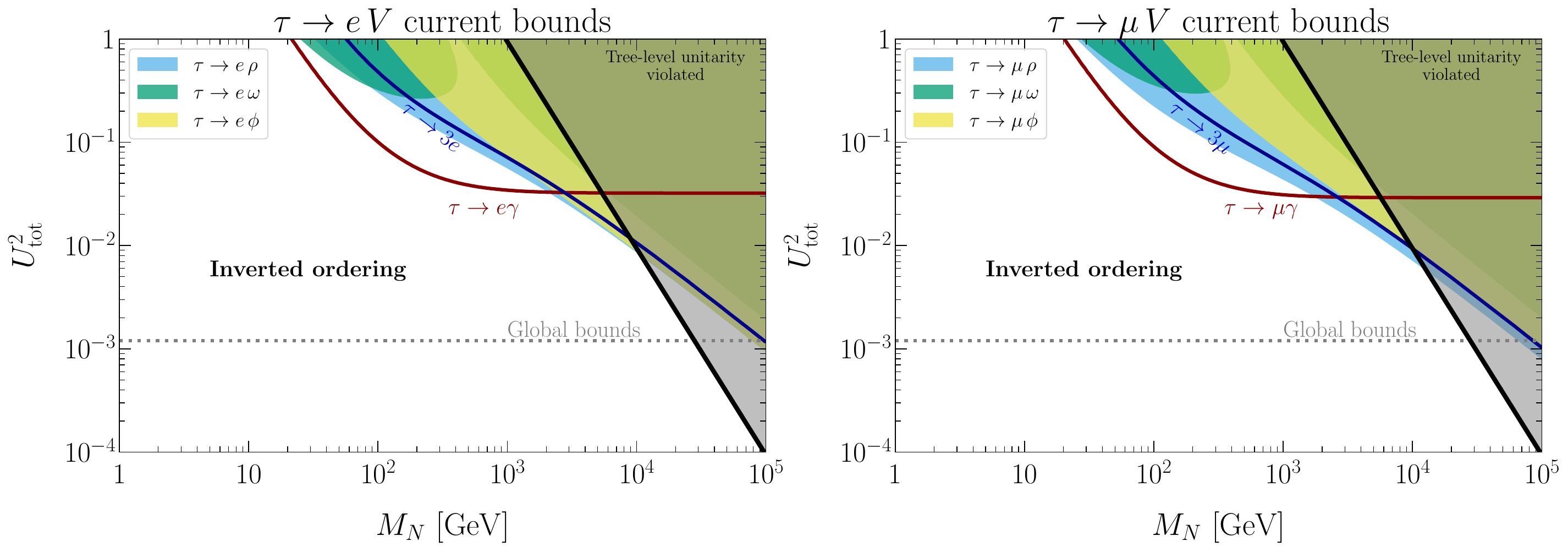}
    \includegraphics[width = \linewidth]{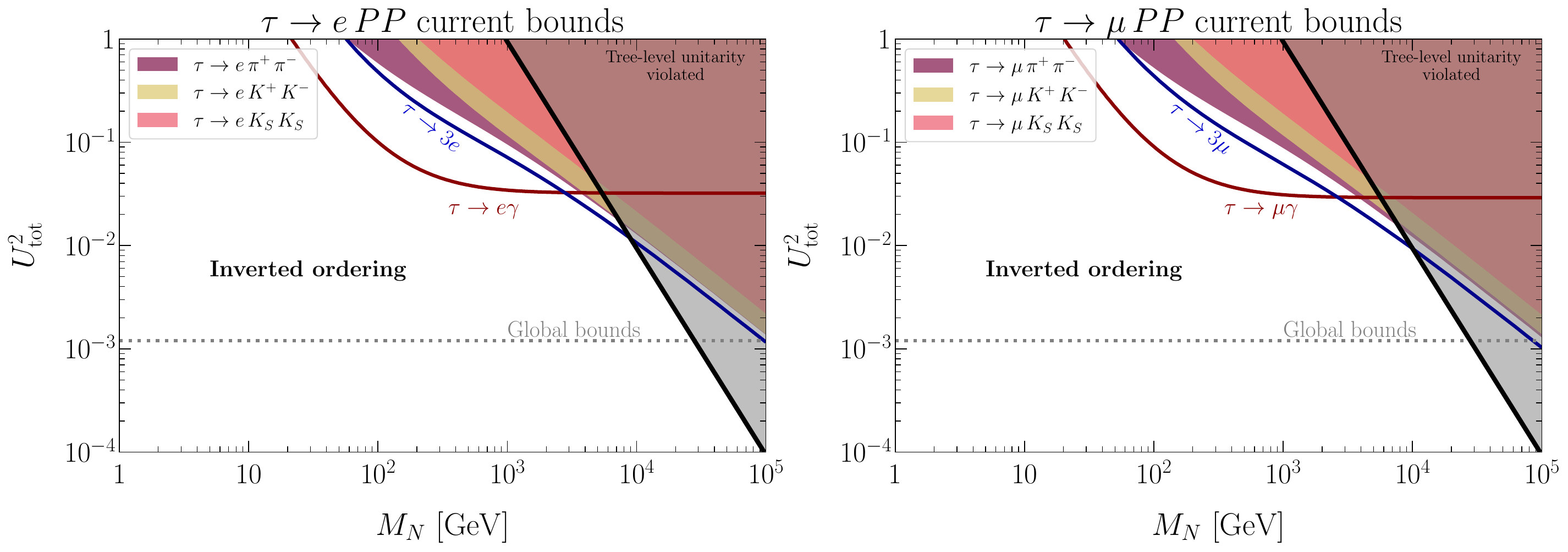}
    \caption{Bounds from $\tau$ cLFV decays with hadrons in the inverted neutrino mass ordering.}
    \label{fig:inverted_ordering_bounds}
\end{figure}

\begin{figure}[!h]
    \centering
    \includegraphics[width = \linewidth]{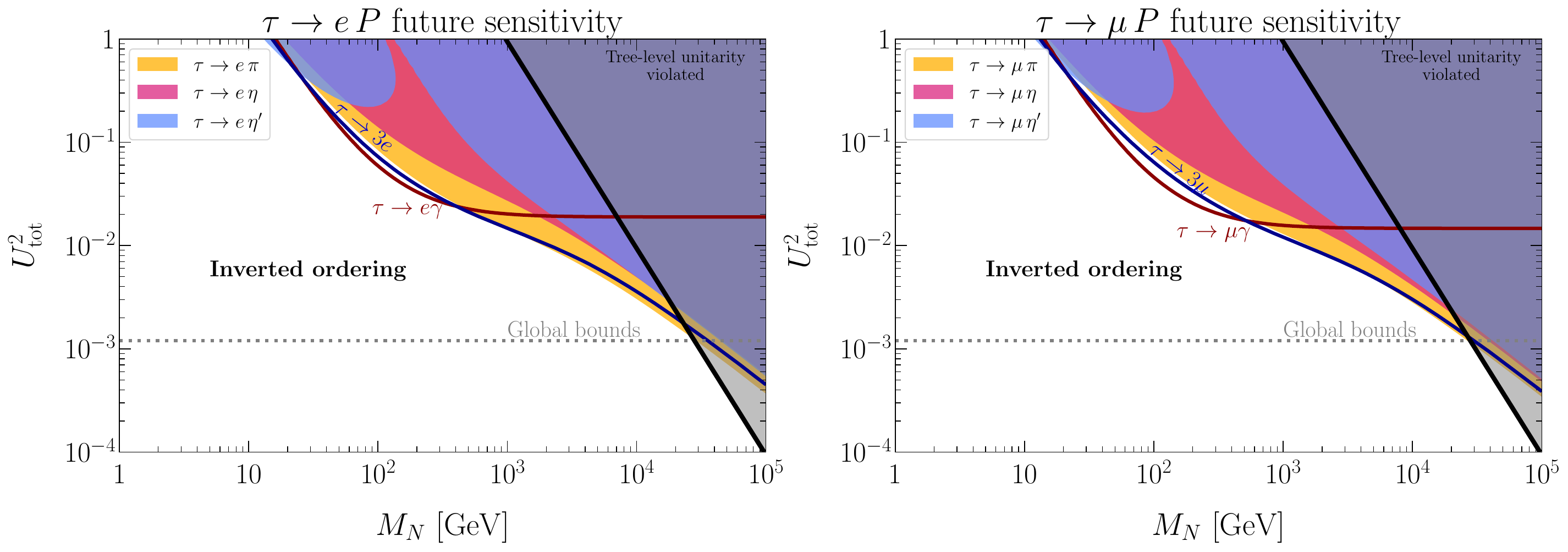}
    \includegraphics[width = \linewidth]{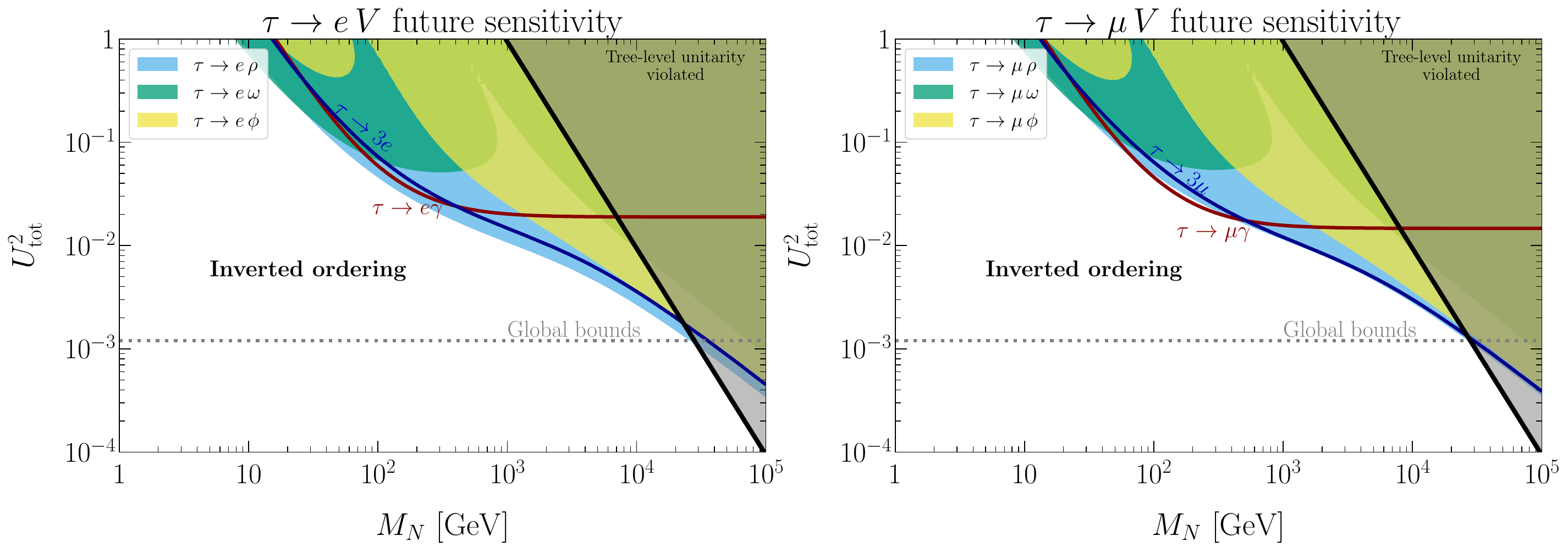}
    \includegraphics[width = \linewidth]{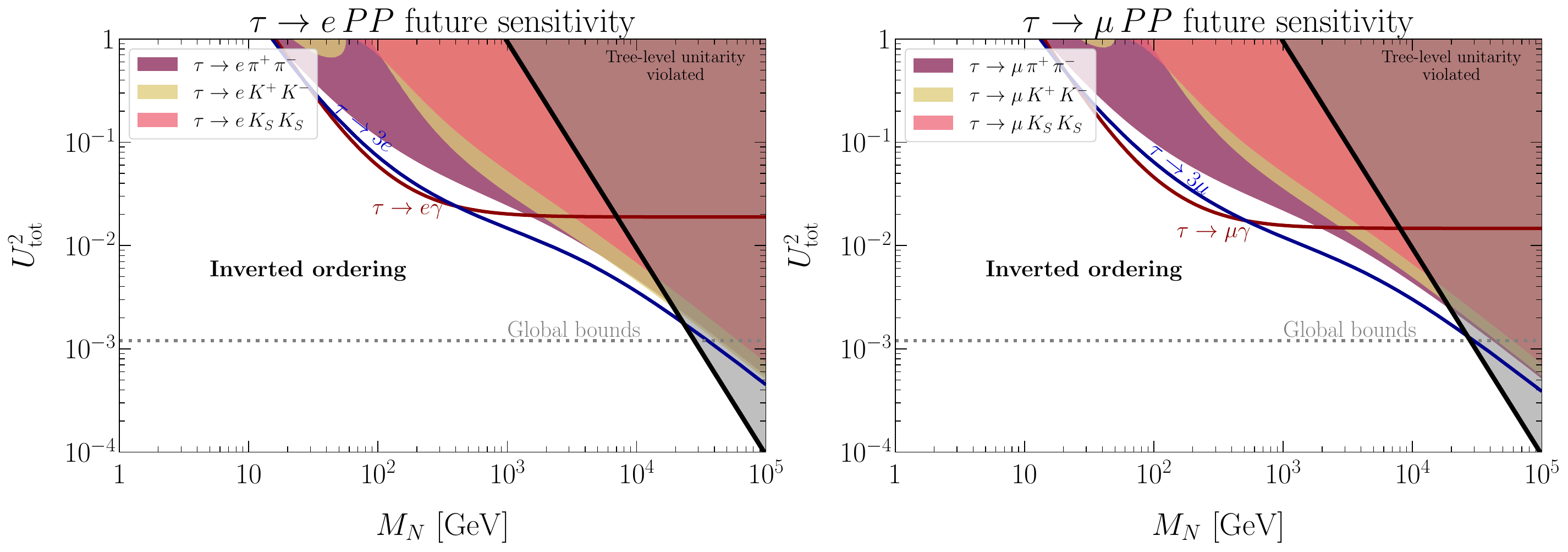}
    \caption{Prospective sensitivity from $\tau$ cLFV decays with hadrons in the inverted neutrino mass ordering.}
    \label{fig:inverted_ordering_future}
\end{figure}

\clearpage
\newpage
\bibliographystyle{JHEP}
\bibliography{bibliography.bib}

\end{document}